\documentclass[11pt,a4paper]{article}
\usepackage[left=2.35cm,top=3cm,bottom=3cm,right=2.3cm]{geometry}

\usepackage{graphicx}
\usepackage[svgnames,x11names]{xcolor}
\definecolor{lightgray}{HTML}{D5D5D5}
\definecolor{lightergray}{HTML}{EEEEEE}
\usepackage{upgreek}
\usepackage{cite}
\usepackage{xcolor,colortbl}

\usepackage[colorlinks=true
,urlcolor=blue
,anchorcolor=blue
,citecolor=blue
,filecolor=blue
,linkcolor=blue
,menucolor=blue
,linktocpage=true
,pdfproducer=medialab
]{hyperref}

\usepackage{amsmath}
\usepackage{amsfonts}
\usepackage{amssymb}
\usepackage{slashed}
\usepackage[caption=false]{subfig}

\usepackage{listings}
\usepackage{pgf}

\newcommand{\D}{\mathrm{d}}

\usepackage{tikz}
\usetikzlibrary{decorations.pathmorphing,positioning,decorations.pathreplacing,shapes,calc}
\usetikzlibrary{decorations.pathreplacing}
\usetikzlibrary{decorations.markings}
\usetikzlibrary{arrows}

\usepackage{enumitem}
\makeatletter
    \newlist{treelist}{itemize}{5}
    \setlist[treelist]{label=\treelist@label}

    \tikzset{treelist line/.style={thick, line cap=round, rounded corners}}
    \def\treelist@label{%
        \begin{tikzpicture}[remember picture, baseline={([yshift=-.6ex] treelist-bullet-\the\enit@depth.center)}]
            \draw [treelist line] (0, 0) -- node (treelist-bullet-\the\enit@depth) {} ++(.5em, 0);
        \end{tikzpicture}%
        \ifnum\enit@depth>1
            \tikz[remember picture, overlay] \draw [treelist line] (treelist-bullet-\the\numexpr\enit@depth-1\relax.center) |- (treelist-bullet-\the\enit@depth.center);%
        \fi
    }
\makeatother

\newcommand{\term}[1]{\emph{#1}}
\newcommand\mcmule{{\sc McMule}}

\def\D{\mathrm{d}}

\def\MeV{{\rm MeV}}
\def\GeV{{\rm GeV}}
\newcommand{\cO}{\mathcal{O}}

\newcommand{\M}[2]{\mathcal{M}_{#1}^{(#2)}}
\newcommand{\fM}[2]{\mathcal{M}_{#1}^{(#2)f}}

\newcommand\ieik{\hat{\mathcal{E}}}
\def\xc{\xi_{c}} 
\newcommand{\cdis}[2][c]{\left(\frac{1}{#2}\right)_{\hspace*{-3pt}#1}}

\begin{document}
\thispagestyle{empty}

\begin{flushright}
PSI-PR-20-09\\
ZU-TH 23/20\\
\end{flushright}
\vspace{3em}
\begin{center}
{\Large\bf QED at NNLO with \mcmule{}}
\\
\vspace{3em}
{\sc
P.\,Banerjee$^{a,}$,
T.\,Engel$^{a,b}$,
A.\,Signer$^{a,b}$,
Y.\,Ulrich$^{a,b}$
}\\[2em]
{\sl ${}^a$ Paul Scherrer Institut,\\
CH-5232 Villigen PSI, Switzerland \\
\vspace{0.3cm}
${}^b$ Physik-Institut, Universit\"at Z\"urich, \\
Winterthurerstrasse 190,
CH-8057 Z\"urich, Switzerland}
\setcounter{footnote}{0}
\end{center}
\vspace{6ex}

\begin{center}
\begin{minipage}{15.3truecm}
{\mcmule{} is a framework for fully differential higher-order QED
calculations of scattering and decay processes involving leptons. It
keeps finite lepton masses, which regularises collinear singularities.
Soft singularities are treated with dimensional regularisation and
using FKS$^\ell$ subtraction. We describe the implementation of the
framework in Fortran~95, list the processes that are currently
implemented, and give instructions on how to run the code. In
addition, we present new phenomenological results for muon-electron
scattering and lepton-proton scattering, including the dominant NNLO
corrections. While the applications presented focus on MUonE, MUSE,
and P2, the code can be used for a large number of planned and running
experiments.}
\end{minipage}
\end{center}

\newpage


\section {Introduction}

Perturbation theory is a well-established tool to provide accurate
theoretical descriptions of many scattering and decay processes. In
fact, it is often the case that the coupling (either electromagnetic,
strong, or electroweak) is small enough to facilitate a perturbative
treatment and non-perturbative effects are either subdominant or can
be isolated and modelled to a sufficient precision. Hence, there has
been a huge effort and impressive progress in computational techniques
for higher-order perturbative calculations.

While most of the effort of the community is geared towards
high-energy colliders, there is also a very important low-energy
programme ongoing. For example elastic electron-proton scattering at
the Jefferson Laboratory lead to a determination of the weak charge of
the proton by QWeak~\cite{Androic:2018kni} or allowed
PRad~\cite{Xiong:2019umf} to provide crucial input towards the
solution of the proton radius puzzle~\cite{Pohl:2010zza,
Antognini:1900ns}. The same process has been measured at MAMI by the
A1 collaboration~\cite{Bernauer:2013tpr} to determine form factors and
will be studied again at MESA, where P2~\cite{Becker:2018ggl} aims at
a precise determination of the weak mixing angle through an asymmetry
measurement at a small beam energy of 155\,\MeV. A similar approach
but using electron-electron scattering is pursued by the Moller
experiment~\cite{Benesch:2014bas}.

Two planned experiments for which we provide new results are
MUonE~\cite{LoI} and MUSE~\cite{Gilman:2017hdr}. The idea of MUonE is
to use a 150\,\GeV\ muon beam at CERN to measure the differential
cross section for elastic muon-electron scattering at a centre-of-mass
energy of $\sqrt{s}\sim 400\,\MeV$. This is motivated by the
connection~\cite{Keshavarzi:2020bfy} of hadronic vacuum polarisation
(HVP) effects with the anomalous magnetic moment of leptons. From the
shape of the muon-electron cross section it is possible to extract the
effective electromagnetic coupling and, hence, to obtain an
independent determination of the leading hadronic contribution. The
idea of MUSE is to measure simultaneously electron-proton and
muon-proton scattering, for positively and negatively charged leptons.
The experiment will be carried out at the Paul Scherrer Institut with
lepton momenta $\cO(100\,\MeV)$ and will shed further light on the
proton radius puzzle and two-photon exchange contributions.

Bhabha scattering is a further example which has been studied
extensively~\cite{Actis:2010gg} in connection with luminosity
measurements. Finally, we mention muon and tau decay processes that
can be described through QED corrections in the Fermi theory.  This
list is tailored towards the applications discussed in this paper and
is by no means complete. But it shows that there is a demand for
precise higher-order QED calculations for low-energy scattering and
decay processes involving leptons. It is the aim of \mcmule{} (Monte
Carlo for MUons and other LEptons) to provide a Monte Carlo code that
can be used to obtain precise theoretical predictions for a wide range
of low-energy processes dominated by QED effects, with a particular
focus on processes involving muons. More precisely, \mcmule{} is an
integrator that allows to obtain histograms for arbitrary, fully
differential observables.

QED calculations are typically simpler than QCD computations. First,
due to the abelian nature of QED, the algebra is less involved.  A
more important aspect is the simplified structure of infrared
singularities in QED, which reduces the complexity of the divergent
phase-space integrations. Generally, it is a highly non-trivial
problem to move from matrix elements to fully differential physical
observables.  However, the abelian gauge structure of QED leads to a
simple Yennie-Frautschi-Suura (YFS) exponentiation of multiple soft
singularities~\cite{Yennie:1961ad}. Also, in QED collinear
singularities are only possible if a gauge boson (photon) becomes
collinear to a fermion. These singularities can be regularised through
non-vanishing fermion masses. 

The relative simplicity of QED might well be responsible for a
remarkable divide in the computational techniques that are used in the
QED and QCD community. Typically, scattering processes in QED are
computed using an infinitesimal photon mass to regularise infrared
singularities and using a slicing method to extract the
infrared-divergent part of phase-space integrations.  In \mcmule{} we
follow more closely techniques familiar from QCD calculations and
therefore use dimensional regularisation and a subtraction scheme. In
this context, the simplicity of the infrared structure of QED has been
exploited in~\cite{Engel:2019nfw}, where a subtraction scheme at
next-to-next-to leading order (NNLO) and beyond has been developed
that allows to obtain arbitrary fully exclusive quantities as soon as
the matrix elements are known.

Despite the simplicity of QED, there is one aspect in which QED
computations are more complicated than QCD calculations. It is related
to potentially large logarithms $\log(m^2/Q^2)$ that are remnants of
collinear singularities. Here, $Q$ is a typical scale of the process,
which is often much larger than some of the fermion masses $m$. In
QCD, quantities are usually considered that are inclusive enough such
that final-state collinear singularities cancel. Hence, no
corresponding large logarithms appear in the final result.
Initial-state collinear singularities are factorised into parton
distribution functions. Thus, it is possible to set $m=0$, often to a
very good approximation. In QED, this is not the case. Many
distributions that are measured are dominated by these logarithms,
such that it is often not possible to work with massless leptons. The
dominance of the logarithmic terms can be exploited to obtain
approximate expressions for higher-order corrections, see e.g.
\cite{Arbuzov:2019hcg} for a review.  Keeping finite fermion masses is
a substantial complication for the evaluation of virtual corrections.
In addition it potentially leads to numerical problems if a fully
differential Monte Carlo approach is taken. Thus, in many cases QED
results cannot be simply extracted from corresponding QCD results, but
a dedicated effort is required. 

The Fortran~95 code \mcmule{} can be downloaded at
\begin{lstlisting}[language=bash]
    (*@\url{https://gitlab.com/mule-tools/mcmule}@*)
\end{lstlisting}
where also an up-to-date table of implemented processes, a
documentation, and some sample results can be found. At the time of
writing, the following processes are implemented:
\begin{align}
    \label{list:processes}
    \ell&\to\ell' \nu \bar{\nu} & & \mbox{NNLO} & \nonumber \\
    \ell&\to\ell' \nu \bar{\nu} \gamma & & \mbox{NLO} &  \nonumber \\
    \ell&\to\ell' \nu \bar{\nu} (l^+ l^-) & & \mbox{NLO} &\\ 
    \ell p &\to \ell p & & \mbox{NLO and dominant NNLO} & \nonumber \\
    \ell \ell' &\to \ell \ell' & & \mbox{NLO and dominant NNLO} &  \nonumber \,
\end{align}
where $\ell$ and $\ell'$ are different leptons and $l$ is either equal
to $\ell'$ or the third possible lepton. The lepton decay processes
are computed in the Fermi theory. For the processes with a proton $p$
the approximation is made whereby its interaction is only due to the
exchange of a single photon.

In this article we will start in Section~\ref{sec:qedcalc} by briefly
recapitulating the techniques we use to do fully differential
higher-order QED calculations. The structure of the code, which
consists of several modules with a simple, mostly hierarchic structure
is described in Section~\ref{sec:structure}. In
Section~\ref{sec:example} we perform a basic leading-order (LO)
calculation in order to illustrate how to run the code. The following
two sections are devoted to our main new phenomenological results. We
start with MUonE in Section~\ref{sec:muone}. First we explain how to
use \mcmule{} to reproduce next-to-leading order (NLO) results
available in the literature~\cite{Alacevich:2018vez}. Then we present
new results for $\mu$-$e$ scattering, including numerically dominant
NNLO corrections. Section~\ref{sec:muse} is devoted to lepton-proton
scattering. We discuss how to extend the partial NNLO calculation of
the previous section to elastic $e$-$p$ and $\mu$-$p$ scattering and
provide some phenomenological results adapted to P2 and MUSE. These
processes are just the beginning of the \mcmule{} programme. In
Section~\ref{sec:future} we discuss possible future developments of
\mcmule{}. Finally, the input parameters used by \mcmule{} are listed
in Appendix~\ref{sec:appendix}.


\section{QED corrections as implemented in \mcmule{}}
\label{sec:qedcalc} 

As mentioned in the introduction, some of the techniques used within
\mcmule{} are somewhat different to what is typically used for
higher-order QED calculations. For a start, infrared singularities due
to soft photons are regularised through dimensional regularisation in
$d=4-2\epsilon$ dimensions. The photon is kept strictly massless also
in intermediate steps.  However, the masses of the fermions are always
kept at their physical value and not set to zero. This regularises all
collinear singularities in QED and gives rise to terms involving the
logarithm of the fermion mass, $\log(m^2/Q^2)$. Such terms often form
the dominant corrections in QED and, thus, it is essential to keep
fermion masses different from zero. This leads to a substantial
complication in the evaluation of virtual corrections. If the two-loop
amplitudes are available only for massless fermions,
massification~\cite{Penin:2005eh, Penin:2005kf, Mitov:2006xs,
Becher:2007cu, Engel:2018fsb} can be used. This is a procedure that
allows to obtain the leading mass terms from the corresponding
massless amplitudes. Two-loop amplitudes obtained this way are
suitable for our approach, but result in the neglect of usually very
small power suppressed terms. While it is possible to partially resum
logarithmic terms, at the current stage no effort is made within
\mcmule{} to do so. Presently, \mcmule{} is a strict fixed-order fully
differential particle/parton-level Monte Carlo integrator.

Since we are dealing with low-energy processes we always renormalise
the fermion masses and coupling in the on-shell scheme. The treatment
of infrared singularities that occur when combining real and virtual
corrections is coded according to the FKS subtraction
method~\cite{Frixione:1995ms, Frederix:2009yq} and its generalisation
beyond NLO for massive QED, FKS$^\ell$~\cite{Engel:2019nfw}.

The core idea of this method is to render the phase-space integration
of a real matrix element finite by subtracting all possible soft
limits. The subtracted pieces are partially integrated over the phase
space and combined with the virtual matrix elements to form finite
integrands.  For a detailed discussion of the method we refer to
\cite{Engel:2019nfw}. Here, we just give a schematic overview with the
basic information required to understand the structure of the code.

The NLO corrections $\sigma^{(1)}$ to a cross section are split into a
$n$-particle and $(n+1)$-particle contribution and are written as
\begin{subequations}
\label{eq:nlo:4d}
\begin{align}
\sigma^{(1)} &=
\sigma^{(1)}_n(\xc) + \sigma^{(1)}_{n+1}(\xc) \, , \\
\sigma^{(1)}_n(\xc) &= \int
\ \D\Phi_n^{d=4}\,\Bigg(
    \M n1
   +\ieik(\xc)\,\M n0
\Bigg) = 
\int \ \D\Phi_n^{d=4}\, \fM n1(\xc)
\,,
\label{eq:nlo:n}
\\
\sigma^{(1)}_{n+1}(\xc) &= \int 
\ \D\Phi^{d=4}_{n+1}
  \cdis{\xi_1} \big(\xi_1\, \fM{n+1}0 \big)
\label{eq:nlo:n1}
\, .
\end{align}
\end{subequations}
In \eqref{eq:nlo:n1}, $\xi_1$ is a variable of the $(n+1)$-parton
phase space $\D\Phi^{d=4}_{n+1}$ that corresponds to the (scaled)
energy of the emitted photon. For $\xi_1\to 0$ the real matrix element
(or more precisely the absolute value squared of the amplitude)
$\fM{n+1}0$ develops a singularity. The superscripts $(0)$ and $f$
indicate that the matrix element is computed at tree level and is
finite, i.e. free of explicit infrared poles $1/\epsilon$. In order to
avoid an implicit infrared pole upon integration, the $\xi_1$
integration is modified by the factor $\xi_1 (1/\xi_1)_c$, where the
distribution $(1/\xi_1)_c$ acts on a test function $f(\xi_1)$ as
\begin{align}
  \label{subcond}
\int_0^1\D\xi_1\, \cdis{\xi_1}\, f(\xi_1)
&\equiv
\int_0^1\D\xi_1\,\frac{f(\xi_1)-f(0)\theta(\xc-\xi_1)}{\xi_1}
\,.
\end{align}
Thus, for $\xi_1 < \xc$, the integrand is modified through the
subtraction of the soft limit $f(0)$. This renders the integration
finite. However, it also modifies the result. The missing piece of the
real corrections can be trivially integrated over $\xi_1$. This
results in the integrated eikonal factor $\ieik(\xc)$ times the
tree-level matrix element for the $n$-particle process, $\M n0$. The
factor $\ieik(\xc)$ has an explicit $1/\epsilon$ pole that cancels
precisely the corresponding pole in the virtual matrix element $\M
n1$. Thus, the combined integrand of \eqref{eq:nlo:n} is free of
explicit poles, hence denoted by $\fM n1$, and can be integrated
numerically over the $n$-particle phase space $\D\Phi_n^{d=4}$. 

The parameter $\xc$ that has been introduced to split the real
corrections can be chosen arbitrarily as long as
\begin{align}
  \label{eq:xic}
0<\xc\le\xi_\text{max} = 1-\frac{\big(\sum_i m_i\big)^2}{s}\,,
\end{align}
where the sum is over all masses in the final state. The $\xc$
dependence has to cancel exactly between \eqref{eq:nlo:n} and
\eqref{eq:nlo:n1} since at no point any approximation was made in the
integration. Checking this independence is a very useful tool to test
the implementation of the method as well as its numerical stability.

The finite matrix element $\fM n1$ is simply the first-order expansion
of the general YFS exponentiation formula~\cite{Yennie:1961ad} for
soft singularities
\begin{align}
e^{\ieik}\, \sum_{\ell = 0}^\infty \M{n}{\ell} = 
\sum_{\ell = 0}^\infty \fM{n}{\ell}
=  \M n0 +
  \Big(\M n1 +\ieik(\xc)\,\M n0\Big) + \mathcal{O}(\alpha^2)\,,
\label{eq:yfsnew}
\end{align}
where we exploited the implicit factor $\alpha$ in $\ieik$.

As detailed in~\cite{Engel:2019nfw}, for QED with massive fermions
this scheme can be extended to NNLO and, in fact, beyond. The NNLO
corrections are split into three parts
\begin{subequations}
\label{eq:nnlo:4d}
\begin{align}
\begin{split}
\sigma^{(2)}_n(\xc) &= \int
\ \D\Phi_n^{d=4}\,\bigg(
    \M n2
   +\ieik(\xc)\,\M n1
   +\frac1{2!}\M n0 \ieik(\xc)^2
\bigg) = 
\int \ \D\Phi_n^{d=4}\, \fM n2(\xc)
\,,
\end{split}\label{eq:nnlo:n}
\\
\sigma^{(2)}_{n+1}(\xc) &= \int 
\ \D\Phi^{d=4}_{n+1}
  \cdis{\xi_1} \Big(\xi_1\, \fM{n+1}1(\xc)\Big)\label{eq:nnlo:n1}
\,,\\
\sigma^{(2)}_{n+2}(\xc) &= \int
\ \D\Phi_{n+2}^{d=4}
   \cdis{\xi_1}\,
   \cdis{\xi_2}\,
     \Big(\xi_1\xi_2\, \fM{n+2}0\Big) \label{eq:nnlo:n2}\, .
\end{align}
\end{subequations}
Thus we have to evaluate $n$-parton contributions, single-subtracted
$(n+1)$-parton contributions, and double-subtracted $(n+2)$-parton
contributions. This structure will be mirrored in the Fortran code.
The $\xc$ dependence cancels, once all three contributions are taken
into account. An example of this will be shown in
Figure~\ref{fig:xicut}.

The method described above has actually already been used for several
processes. The radiative~\cite{Pruna:2017upz} and rare
decay~\cite{Pruna:2016spf} of the muon and tau~\cite{Ulrich:2017adq}
have been implemented at NLO in the Fermi theory in a fully
differential code. In addition, the Michel decay of the muon has been
added at NNLO~\cite{Engel:2019nfw}. These results have been verified
by comparison to more analytic and more inclusive
computations~\cite{Fael:2015gua, Fael:2016yle, Fael:2019usp,
Anastasiou:2005pn, Pak:2008cp}. Thus, the method is fully established
and \mcmule{} can be seen as a natural extension of these previous
computations and a container to include further phenomenologically
relevant processes.


\section{Structure of \mcmule{}}
\label{sec:structure} 

\mcmule{} is written in Fortran 95 with helper and analysis tools
written in {\tt python}\footnote{Additionally to the {\tt python} tool
a Mathematica tool is available.}. An online documentation can be
found at the git repository listed in the
introduction~\cite{mcmuleman}. The code is written with two kinds of
applications in mind. First, several processes are implemented, some
at NLO, some at NNLO. Since new processes are continuously added, we
refer to the online documentation for a list of available processes.
For these, the user can define an arbitrary (infrared safe), fully
differential observable and compute cross sections and distributions.
Second, the program is set up such that additional processes can be
implemented by supplying the relevant matrix elements.

\vbox{To obtain a copy of \mcmule{} we recommend the following approach
\begin{lstlisting}[language=bash]
$ git clone --recursive https://gitlab.com/mule-tools/mcmule
\end{lstlisting}}
\noindent To build \mcmule{}, a Fortran compiler such as {\tt gfortran} and a
python installation is needed. The main executable can be compiled by
running 
\vspace{1mm}
\begin{lstlisting}[language=bash]
$ ./configure
$ make mcmule
\end{lstlisting}
\vspace{1mm}
Alternatively, we provide a Docker container~\cite{Merkel:2014} for
easy deployment and legacy results. In multi-user environments, {\sl
udocker}~\cite{Gomes:2017hct} can be used instead. In either case, a
pre-compiled copy of the code can be obtained by calling
\vspace{1mm}
\begin{lstlisting}[language=bash]
$ docker pull yulrich/mcmule  # requires Docker to be installed
$ udocker pull yulrich/mcmule # requires uDocker to be installed
\end{lstlisting}
\vspace{1mm}
We provide instructions on how \mcmule{} is used in
Section~\ref{sec:example}.

When started, {\tt mcmule} reads options from {\tt stdin} as specified
in Table~\ref{tab:mcmuleinput} of Section~\ref{sec:example}.  The
value and error estimate of the integration is printed to {\tt stdout}
and the full status of the integration is written in a
machine-readable format into a folder called {\tt out/} (see below).

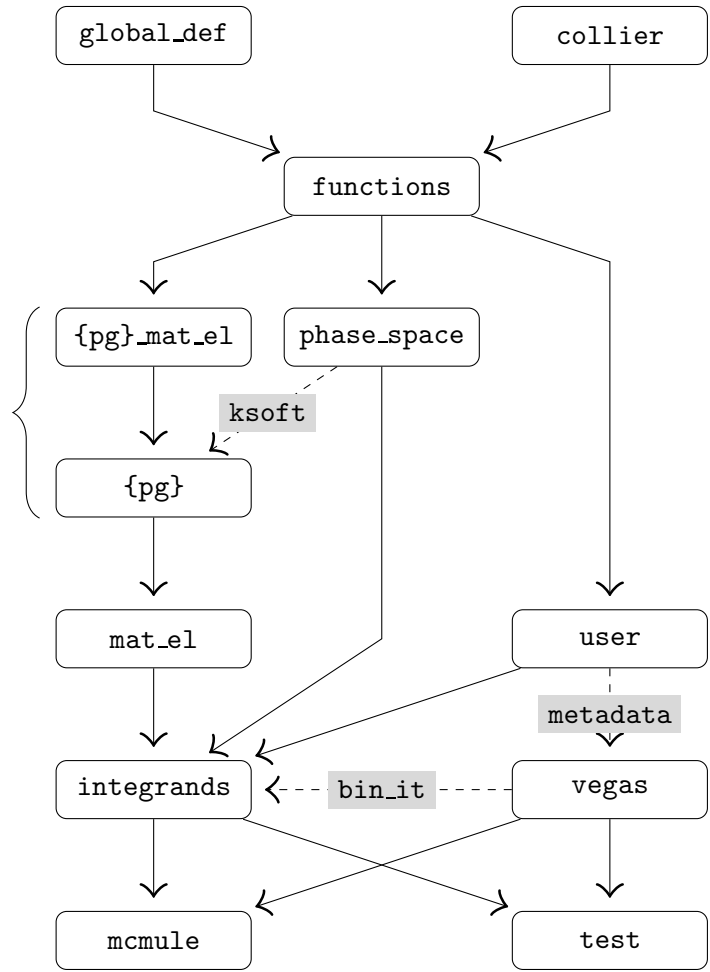
\begin{figure}
  \centering

\tikzset{
    block/.style={
        rectangle, draw, 
        text width=6em, minimum height=2em, 
        text centered, rounded corners
    },
    pblock/.style={
        rectangle, fill=gray!30, midway
    },
    line/.style={
        decoration={markings,mark=at position \pgfdecoratedpathlength-5pt with {\arrow[scale=3]{>}}},
        postaction={decorate}, shorten >= 5pt
    }
}
\begin{tikzpicture}[x=3cm,y=-2cm]
\node[block       ] (GD)  at (-1,-1) {\tt global\_def};
\node[block       ] (CLL) at ( 1,-1) {\tt collier};
\node[block       ] (FU)  at ( 0, 0) {\tt functions};
\node[block       ] (USR) at ( 1, 3) {\tt user};
\node[block       ] (PS)  at ( 0, 1) {\tt phase\_space};
\node[block       ] (ML)  at (-1, 1) {\tt \string{pg\string}\_mat\_el};
\node[block       ] (PG)  at (-1, 2) {\tt \string{pg\string}};
\node[block       ] (MG)  at (-1, 3) {\tt mat\_el};
\node[block       ] (INT) at (-1, 4) {\tt integrands};
\node[block       ] (VEG) at ( 1, 4) {\tt vegas};
\node[block       ] (XS)  at (-1, 5) {\tt mcmule};
\node[block       ] (TST) at ( 1, 5) {\tt test};

\draw [line       ] (CLL) --+ ( 0,0.5) -- (FU) ;
\draw [line       ] (GD)  --+ ( 0,0.5) -- (FU) ;
\draw [line       ] (FU)  --+ ( 1,0.5) -- (USR);
\draw [line       ] (FU)               -- (PS) ;
\draw [line       ] (FU)  --+ (-1,0.5) -- (ML) ;
\draw [line       ] (ML)               -- (PG) ;
\draw [line       ] (PG)               -- (MG) ;
\draw [line       ] (MG)               -- (INT);
\draw [line,dashed] (PS)               -- (PG)   node[pblock] {\tt ksoft};
\draw [line       ] (PS)  --+ (0,2)    -- (INT);
\draw [line       ] (USR)              -- (INT);
\draw [line,dashed] (USR)              -- (VEG) node[pblock] {\tt metadata};
\draw [line,dashed] (VEG)              -- (INT) node[pblock] {\tt bin\_it};
\draw [line       ] (INT)              -- (XS) ;
\draw [line       ] (VEG)              -- (XS) ;
\draw [line       ] (INT)              -- (TST);
\draw [line       ] (VEG)              -- (TST);

\draw [decorate,decoration={brace,amplitude=10pt}] ( -1.5,2.2) --
(-1.5,0.8);
\end{tikzpicture}
  \caption{The structure of \mcmule{}}
  \label{fig:structure}
\end{figure}

\mcmule{} consists of several modules with a simple, mostly hierarchic
structure. The relation between the most important Fortran modules is
depicted in Figure~\ref{fig:structure}. A solid arrow indicates
``using'' the full module, whereas a dashed arrow is indicative of
partial use. In what follows we give a brief description of the
various modules and mention some variables that play a prominent role
in the interplay between the modules.

\begin{description}
    \item[{\tt global\_def}:] 
    This module simply provides some parameters such as fermion masses
    that are needed throughout the code. It also defines {\tt prec} as
    a generic type for the precision used.\footnote{For quad precision
    {\tt prec=16} and the compiler flag {\tt -fdefault-real-16} is
    required.} Currently, this simply corresponds to double precision.

    \item[{\tt functions}:] 
    This is a library of basic functions that are needed at various
    points in the code. This includes dot products, eikonal factors,
    the integrated eikonal, and an interface for scalar integral
    functions among others.

    \item[{\tt collier}:] 
    This is an external module~\cite{Denner:2016kdg, Denner:2010tr,
    Denner:2005nn, Denner:2002ii}. It will be linked to \mcmule{}
    during compilation and provides the numerical evaluations of the
    scalar and in some cases tensor integral functions in {\tt
    functions}.

    \item[{\tt phase\_space}:] 
      The routines for generating phase-space points and their weights
      are collected in this module.  Phase-space routines ending with
      {\tt FKS} are prepared for the FKS subtraction procedure with a
      single unresolved photon. In the weight of such routines a
      factor $\xi_1$ is omitted to allow the implementation of the
      distributions in the FKS method, see \eqref{eq:nlo:n1}. This
      corresponds to a global variable {\tt xiout}. This factor has to
      be included in the integrand of the module {\tt{integrands}}.
      Also the variable {\tt ksoft} is provided that corresponds to
      the photon momentum without the (vanishing) energy factor
      $\xi_1$. Routines ending with {\tt FKSS} are routines with two
      unresolved photons, see \eqref{eq:nnlo:n2}. Correspondingly, a
      factor $\xi_1\,\xi_2$ is missing in the weight. The global
      variables {\tt xiout1} and {\tt xiout2} as well as {\tt ksoft1}
      and {\tt ksoft2} are provided.\footnote{In the current version
      of \mcmule{} these variables are called {\tt xioutA}, {\tt
      xioutB}, {\tt ksoftA}, and {\tt ksoftB}.}

    \item[{\tt \string{pg\string}\_mat\_el}:]
    Matrix elements are grouped into process groups such as muon decay
    ({\tt mudec}) or $\mu$-$e$ and $\mu$-$p$ scattering ({\tt mue}).
    Each process group contains a {\tt mat\_el} module that provides
    all matrix elements for its group.  Simple matrix elements are
    coded directly in this module. More complicated results are
    imported from sub-modules not shown in Figure~\ref{fig:structure}.
    A matrix element starting with {\tt P} contains a polarised
    initial state.  A matrix element ending in {\tt av} is averaged
    over a neutrino pair in the final state.

    \item[{\tt \string{pg\string}}:] 
    In this module the soft limits of all applicable matrix elements
    of a process group are provided to allow for the soft subtractions
    required in the FKS scheme. These limits are simply the eikonal
    factor evaluated with {\tt ksoft} from {\tt phase\_space} times
    the reduced matrix element, provided through {\tt mat\_el}.

    This module also functions as the interface of the process group,
    exposing all necessary functions that are imported by

    \item[{\tt mat\_el},] which collects all matrix elements as well
    as their particle labelling or particle  identification.
    
    \item[{\tt user}:]
    For a user of the code who wants to run for an already implemented
    process, this is the only relevant module.  At the beginning of
    the module, the user has to specify the number of quantities to be
    computed, {\tt nr\_q}, the number of bins in the histogram, {\tt
    nr\_bins}, as well as their lower and upper boundaries, {\tt
    min\_val} and {\tt max\_val}. The last three quantities are arrays
    of length {\tt nr\_q}. The quantities themselves, i.e. the
    measurement function, is to be defined by the user in terms of the
    momenta of the particles in {\tt quant}. Cuts can be applied by
    setting the logical variable {\tt pass\_cut} to
    false\footnote{Technically, {\tt pass\_cut} is a list of length
    {\tt nr\_q}, allowing to decide whether to cut for each histogram
    separately.}. Some auxiliary functions like (pseudo)rapidity,
    transverse momentum etc. are predefined in {\tt functions}. Each
    quantity has to be given a name through the array {\tt names}.

    Further, {\tt user} contains a subroutine called {\tt inituser}.
    This allows the user to read additional input at runtime, for
    example which of multiple cuts should be calculated. It also
    allows the user to print some information on the configuration
    implemented. 

    \item[{\tt vegas}:] 
    As the name suggests this module contains the adaptive Monte Carlo
    routine {\tt vegas}~\cite{Lepage:1980jk}.  The binning routine
    {\tt bin\_it} is also in this module, hence the need for the
    binning metadata, i.e. the number of bins and histograms ({\tt
    nr\_bins} and {\tt nr\_q}, respectively) as well as their bounds
    ({\tt min\_val} and {\tt max\_val}) and names, from {\tt user}.

    \item[{\tt integrands}:] In this module the functions that are to
    be integrated by {\tt vegas} are coded. There are three types of
    integrands: non-subtracted, single-subtracted, and
    double-subtracted integrands, corresponding to the three parts
    of~\eqref{eq:nnlo:4d}. The matrix elements to be evaluated and the
    phase-space routines used are set using function pointers through
    a subroutine {\tt initpiece}. The factors $\xi_i$ that were
    omitted in the phase-space weight have to be included here for the
    single- and double-subtracted integrands.
    
    \item[{\tt mcmule}:]
    This is the main program, but actually does little else than read
    the inputs and call {\tt vegas} with a function provided by {\tt
    integrands}.

    \item[{\tt test}:]
    For developing purposes, a separate main program exists that is
    used to validate the code after each change. Reference values for
    matrix elements and results of short integrations are stored here
    and compared against.

\end{description}

The library of matrix elements deserves a few comments. As matrix
elements quickly become very large, we store them separately from the
main code. This makes it also easy to extend the program by minimising
the code that needs to be changed.

\begin{figure}
\input{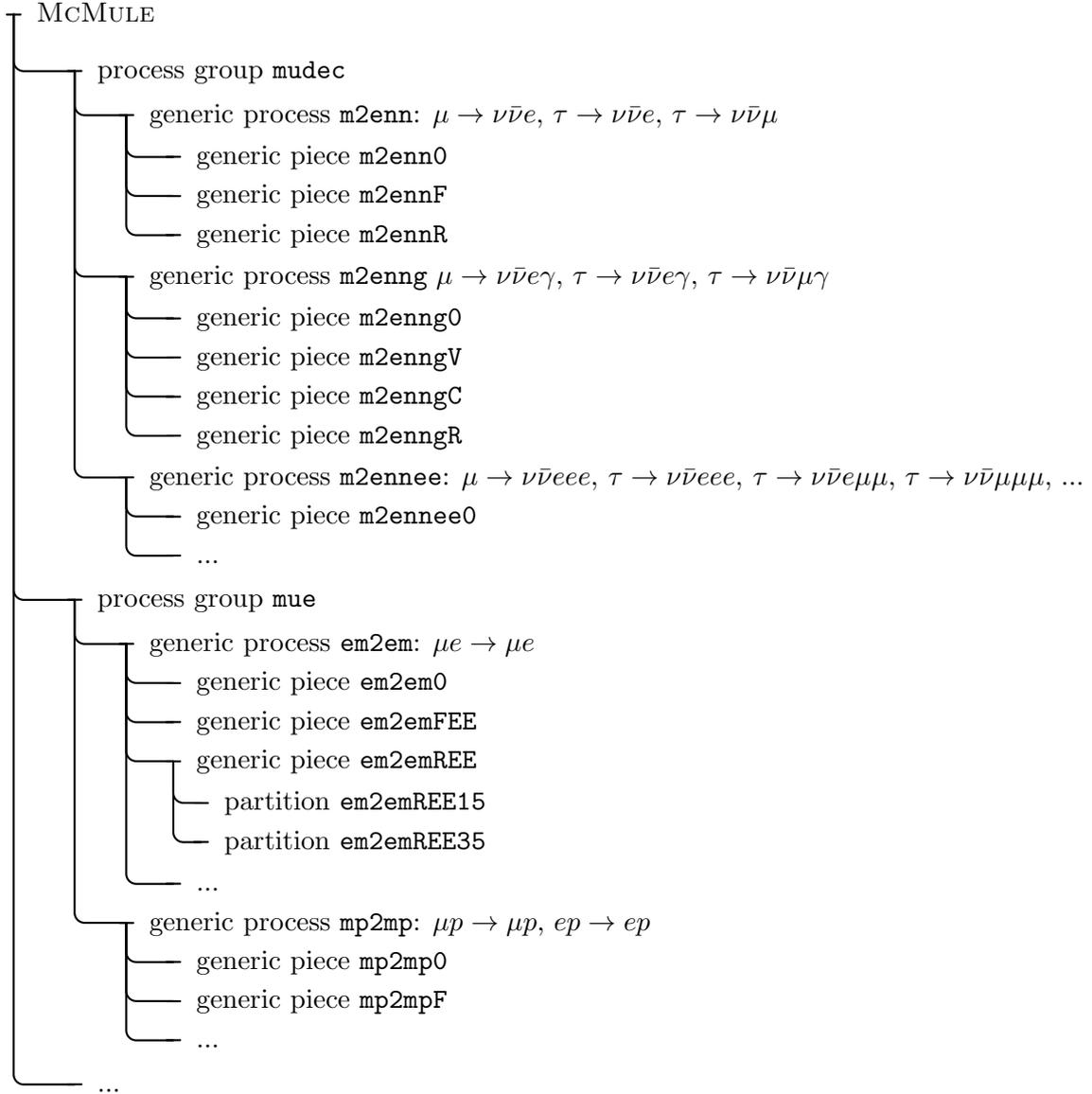}
\caption{The structure of process group, generic process, and generic
piece as used by \mcmule{}. The suffices {\tt 0}, {\tt V}, {\tt C},
{\tt F}, {\tt R}, and others are explained in more detail in
Section~\ref{sec:muone}.}
\label{fig:processtree}
\end{figure}

We group the various contributions into \term{process group},
\term{generic processes}, and \term{generic pieces} as indicated in
Figure~\ref{fig:processtree}. The generic process is a prototype for
the physical process such as $\ell p\to\ell p$ (cf.
Section~\ref{sec:muse}) where the flavour of the lepton $\ell$ is left
open. The generic piece describes a part of the calculation such as
the real or virtual corrections, i.e. the different pieces
of~\eqref{eq:nlo:4d} (or correspondingly~\eqref{eq:nnlo:4d} at NNLO),
that themselves may be further subdivided as is convenient.  In
particular, in some cases a generic piece is split into various
phase-space partitions, as in the example of {\tt em2emREE} in
Figure~\ref{fig:processtree}. A more detailed listing of the various
contributions required for $\mu$-$e$ scattering is given in
Figure~\ref{fig:muonetree}.

When running {\tt mcmule}, the code generates a statefile from which
the full state of the integrator can be reconstructed should the
integration be interrupted. This makes the statefile ideal to also
store results in a compact format.  To analyse these results, we
provide a python tool {\tt pymule}, additionally to the main code for
\mcmule{}. The tool {\tt pymule}, which can be found under {\tt
tools/pymule}, uses {\tt numpy}~\cite{Walt:2011np} for data storage
and {\tt matplotlib} for plotting~\cite{Hunter:2007mp}. While {\tt
pymule} works with any python interpreter, {\tt
IPython}~\cite{Perez:2007ip} is recommended.  A full list of functions
provided can be found in the online manual of {\tt
pymule}~\cite{mcmuleman}.

An important issue are numerical instabilities arising in problematic
regions of the phase space. This is typically the case if an emitted
photon becomes soft or collinear to a massive, but light, fermion. For
soft photon emission the numerical instability is related to the FKS
subtraction discussed in Section~\ref{sec:qedcalc}. When $\xi_1$
becomes very small, the difference $f(\xi_1) - f(0) \theta(\xc-\xi_1)$
in~\eqref{subcond} becomes potentially troublesome as $f(\xi_1)$ can
be calculated less precisely than $f(0)$. To avoid this, we choose a
very small {\tt softcut}, below which we set the integrand directly to
zero. In the collinear case small fermion masses give rise to
pseudo-collinear singularities that further complicate a numerical
stable evaluation of the matrix element. \mcmule{} addresses this
issue through a dedicated tuning of the phase-space parametrisation to
help the {\tt vegas} integration find and deal with these problematic
regions. In addition, a {\tt collcut} is applied if the photon becomes
very collinear to a light fermion. During development, {\tt softcut}
and {\tt collcut} are varied to make sure that, within the integration
error, the cross section is independent of the chosen values.
Afterwards, a suitable value is chosen and hard-coded. However, the
user retains the ability to modify this in {\tt inituser}.

\vspace{3mm}
\section{Running \mcmule{}: double radiative muon decay as an example}
\label{sec:example}

In order to provide a simple example with concrete instructions on how
to run the code and to illustrate how it works, we consider the double
radiative decay of the muon $\mu\to e [\nu\bar\nu] \gamma\gamma $ at
leading order. Since the neutrinos are not detected, we average over
them, indicated by the brackets. Hence, we have to be fully inclusive
with respect to the neutrinos. But the code allows to make any cut on
the other final-state particles. 

To be concrete let us assume we want to compute two distributions, the
missing energy $\slashed{E}\equiv E(\mu)-E(e)-E(\gamma_1)-E(\gamma_2)$
and $\cos\theta_e$, the cosine of the angle between the outgoing
positron and the muon polarisation. Both quantities are determined in
the rest frame of the decaying muon. Of course, $\slashed{E}$
corresponds to the combined energies of the neutrinos. To avoid an
infrared singularity in the branching ratio, we have to require a
minimum energy of the photons. We choose this to be $E_\gamma \ge
10~\MeV$ individually for both photons. In addition, we require for
the angle between the two photons $\theta_{\gamma\gamma} > 15^\circ$.

As mentioned in Section~\ref{sec:structure} the quantities are defined
in the module {\tt user} (file {\tt src/user.f95}). At the beginning
of the module we set
\begin{lstlisting}
  nr_q = 2
  nr_bins = 50
  min_val = (/ 0._prec, -1._prec /)
  min_val = (/ 50._prec, 1._prec /)
\end{lstlisting}
where we have decided to have 50 bins for both distributions and {\tt
nr\_q} determines the number of distributions. The boundaries for the
distributions are set as $0 < \slashed{E} < 50~\MeV$ and $-1 \le
\cos\theta_e \le 1$.

The quantities themselves are defined in the function {\tt quant}.
This function takes arguments {\tt q1} to {\tt q7}. These are the
momenta of the particles, arrays of length 4 with the fourth entry the
energy. To figure out which momentum corresponds to which particle the
user needs to check the headers in the module {\tt mat\_el} or in the
manual~\cite{mcmuleman}. In our case, we find
\begin{lstlisting}
  !! From file mudec_pm2ennggav.f95
  use mudec, only: pm2ennggav!!(p1, n1, p2, p3, p4, p5, p6)
    !! mu+(p1) -> e+(p2) \nu_e \bar{\nu}_\mu g(p5) g(p6)
    !! mu-(p1) -> e-(p2) \bar{nu}_e  \nu_\mu g(p5) g(p6)
    !! for massive (and massless) electron
    !! average over neutrino tensor taken
\end{lstlisting}
Indicating that we have {\tt p1} for the incoming $\mu$, {\tt p2} for
the outgoing $e$, and {\tt p5} and {\tt p6} for the two outgoing
photons.  The momenta of the neutrinos must be given but do not enter,
as we average over them.

Schematically, the function {\tt quant} might look like
\begin{lstlisting}
FUNCTION QUANT(P1,P2,P3,P4,P5,P6,P7)
.
.
pass_cut = .true.
pol1 = (/ 0._prec, 0._prec, 0.85_prec, 0._prec /)

ez = (/ 0._prec, 0._prec, 1._prec, 0._prec /)

if(p5(4) < 10._prec .or. p6(4) < 10._prec) pass_cut = .false.
if(cos_th(p5,p6) > 0.965926) pass_cut = .false.

Emiss = p1(4)-p2(4)-p5(4)-p6(4)
names(1) = 'Emiss'
quant(1) = emiss
names(2) = 'CangE'
quant(2) = cos_th(p2,ez)

END FUNCTION QUANT
\end{lstlisting}
Here, we have used the function {\tt cos\_th} provided by the module
{\tt functions}. This returns the cosine of the angle between the two
momenta given as arguments. We have also specified the polarisation
vector {\tt pol1} in accordance with the $\mu^+$ beam used by MEG.
This polarisation has been measured~\cite{Baldini:2015lwl} to be
$P_\mu = -0.85\pm 0.05$. Since \mcmule{} defines the polarisation
through $\mu^-$, the sign has to be changed.  The variable {\tt
pass\_cut} controls the cuts. Initially it is set to true, to indicate
that the event is kept.  Applying a cut amounts to setting {\tt
pass\_cut} to false. 

All that remains to be done is to prepare the input read by {\tt
mcmule} from {\tt stdin}, as specified in Table~\ref{tab:mcmuleinput}.

\begin{table}[t]
\centering
\begin{tabular}{l|l|l}
\bf Variable name& \bf Data type  & \bf Comment
 \\\hline

\tt nenter\_ad   & \tt integer    & calls / iteration during pre-conditioning  \\
\tt itmx\_ad     & \tt integer    & iterations during pre-conditioning         \\
\tt nenter       & \tt integer    & calls / iteration during main run          \\
\tt itmx         & \tt integer    & iterations during main run                 \\
\tt ran\_seed    & \tt integer    & random seed                          \\
\tt xinormcut1   & \tt real(prec) & the $0<\xc\le1$ parameter                  \\
\tt xinormcut2   & \tt real(prec) & the second $\xc$ parameter for
NNLO (or the $\delta_{\text{cut}}$)                                               \\
\tt which\_piece & \tt char(20)   & the part of the calculation to perform     \\
\tt flavour      & \tt char(8)    & the particles involved                     \\
(opt)            & unknown        & the user can request further input during 
                                    {\tt userinit}
\end{tabular}
\caption{The options read from {\tt stdin} by \mcmule{}. The calls are
multiplied by 1000.}
\label{tab:mcmuleinput}
\end{table}

To be concrete let us assume we want to use 10 iterations with
$1000\times 10^3$ points each for pre-conditioning and 20 iterations
with $5000\times 10^3$ points each for the actual numerical
evaluation. We pick a random seed, say 24225, and for the input
variable {\tt which\_piece} we enter {\tt m2enngR}. Since the double
radiative muon decay is not on the list of
processes~\eqref{list:processes}, we actually compute the real
corrections (hence the suffix {\tt R}) of the generic process $\mu\to
e \nu\bar\nu\gamma$. The {\tt flavour} variable is set to {\tt mu-e}.
We could e.g. use {\tt tau-e} to change from the generic process
$\mu\to e \nu\bar\nu\gamma$ to the process $\tau\to e
\nu\bar\nu\gamma$.  This system will be used for other processes as
well. The input variable {\tt which\_piece} determines the generic
process and the part of it that is to be computed (i.e. tree level,
real, double-virtual etc.). In a second step, the input {\tt flavour}
associates actual numbers to the parameters entering the matrix
elements and phase-space generation.

Thus, we run the code by giving the input
\vspace{2mm}
\begin{lstlisting}[language=bash]
$ ./mcmule
1000
10
5000
20
24225
0.1
0.1
m2ennR
mu-e
\end{lstlisting}
\vspace{2mm}
In practice the input will typically not be given by hand. We mention
a more efficient way in Section~\ref{sec:muone} as well as the
manual~\cite{mcmuleman}. The two variables {\tt xinormcut1} and {\tt
xinormcut2} have no effect at all for a tree-level calculation and
will be discussed in Section~\ref{sec:muoneRun} in the context of the
NLO and NNLO run for muon-electron scattering. We also ignore the
optional input for the moment.

Alternatively, \mcmule{} can be run using Docker or {\sl udocker}
without compiling it first by running
\vspace{2mm}
\begin{lstlisting}[language=bash]
$ ./tools/run-docker.sh -i yulrich/mcmule:latest \
     -u path/to/user.f95  -r
\end{lstlisting}
\vspace{2mm}
followed by the same input as above.

\begin{figure}[t]
\centering
\begin{tikzpicture}
\node at (-4cm,0) {\includegraphics[scale=0.45]{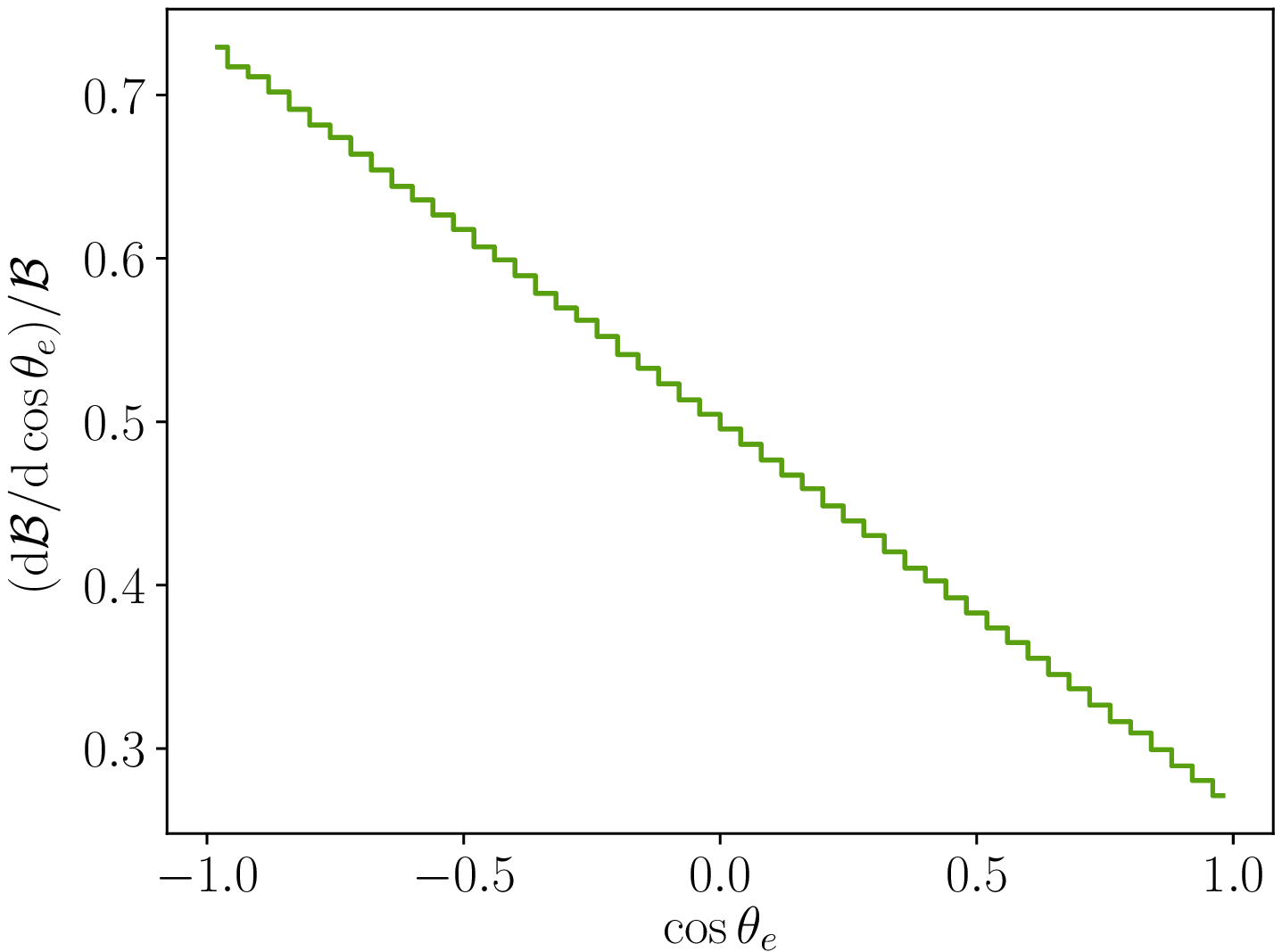}};
\node at (4cm,-3.4mm) {\includegraphics[scale=0.45]{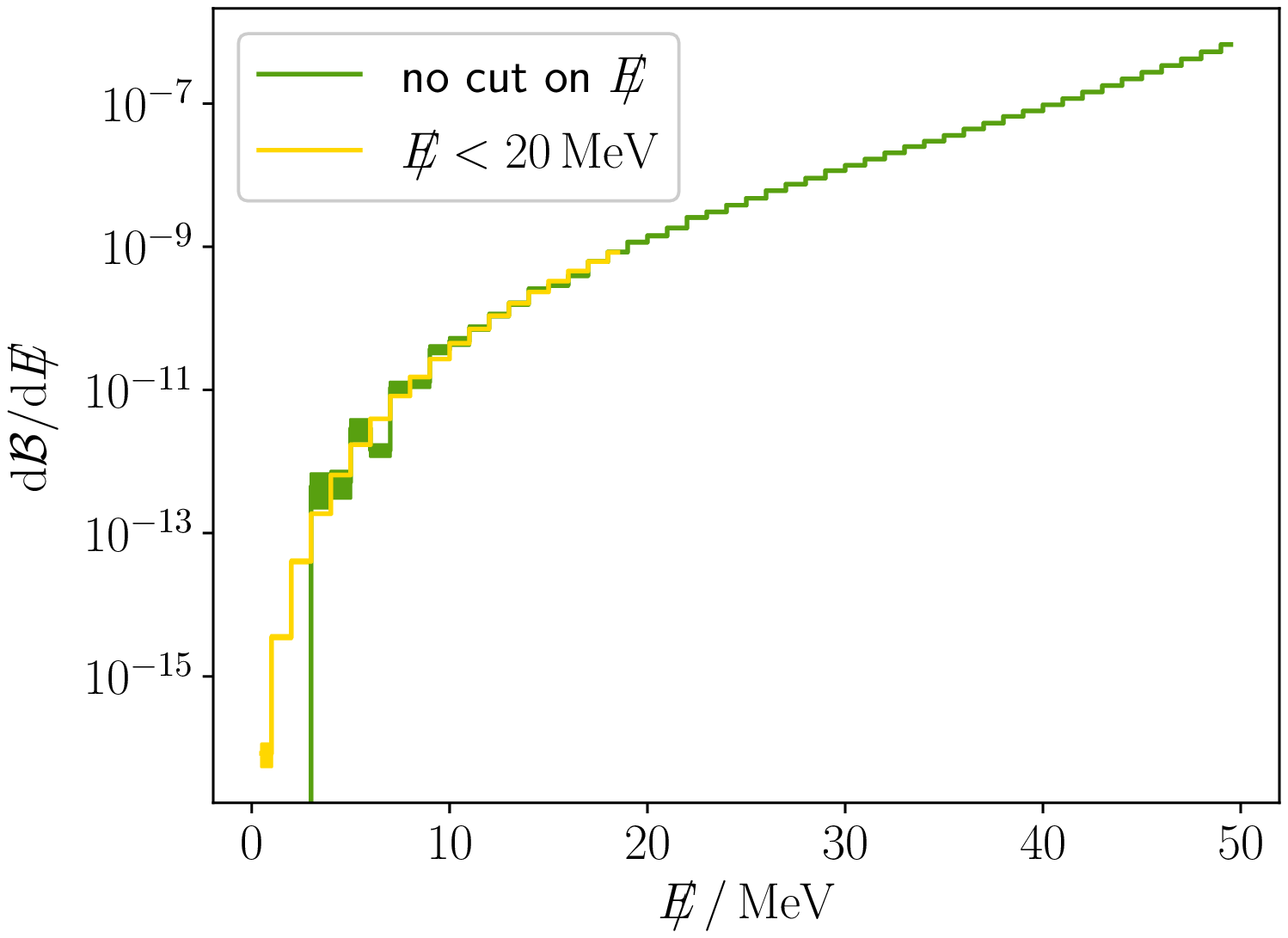}};
\end{tikzpicture}
\caption{Results of a short test run for the branching ratio at LO for
  the double radiative muon decay $\mu\to e \nu\bar{\nu}
  \gamma\gamma$, as a function of the missing energy and the angle of
  the outgoing positron. For the region $\slashed{E}<20\,\MeV$ a
  tailored run is shown in yellow.}
\label{fig:toydrmd}
\end{figure}

Now the mule is ready to trot. After about fifteen minutes on an Intel
i5 processor, it returns the total cross section as\footnote{Unless
otherwise stated, all numerical results have been obtained by running
the code in Docker or {\sl udocker}.}
\begin{lstlisting}
 result, error:  {  1.10052E+06,  1.80838E+02 };   chisq:  0.83
\end{lstlisting}
which, after adding the factor $G_F^2 \alpha^2/\Gamma_\mu$ results in
a branching ratio of $\mathcal{B} = 2.6611(4)\cdot 10^{-5}$. Here
$\Gamma_\mu$ is the measured width of the muon and $G_F$ the Fermi
constant as given in the Appendix~\ref{sec:appendix}. The error is of
course only the statistical error of the Monte Carlo and not a theory
error.  In addition, the two distributions are written into a binary
file that contains the full state of the integrator {\tt
out/m2enngR\_mu-e\_S0000024225X1.00000D1.00000\_ITMX020x005M.vegas}.
The corresponding results are shown as green histograms in
Figure~\ref{fig:toydrmd}, where $d\mathcal{B}/d\cos\theta_e$ has been
normalised.

The results have rather poor statistics. In particular the precision
in the low-energy tail of the sharply falling $\slashed{E}$
distribution is very low since the Monte Carlo generates very few
points there. If the user is interested in this tail it is advisable
to perform dedicated runs. This can be done simply by adding a cut
like
\begin{lstlisting}
  if(emiss > 20.) pass_cut = .false.
\end{lstlisting}
in {\tt quant}. The Monte Carlo will then adapt and result in a more
precise determination of the $\slashed{E}$ distribution in the region
$\slashed{E} < 20\,\MeV$. For illustration in Figure~\ref{fig:toydrmd}
such a tailored run with the same statistics is overlayed in yellow to
the original run in the plot for $\slashed{E}$.

This is all that is required for simply running the code. In what
follows we give a brief outline how the code works. The first step it
does in {\tt mcmule} is to associate the numerical values of the
masses, as specified through {\tt flavour}. In particular, the generic
masses {\tt Mm} and {\tt Me} are set to {\tt Mmu} and {\tt Mel}. This
is done in {\tt initflavour(scms)}, defined in {\tt global\_def}. For
other processes this might also involve setting e.g. centre-of-mass
energies {\tt scms} to default values.

Next, the function to be integrated by {\tt vegas} is determined. This
is a function stored in {\tt integrands}. There are basically three
types of integrands: a standard, non-subtracted integrand {\tt
sigma\_0}, a single-subtracted integrand {\tt sigma\_1} needed beyond
LO, and a double-subtracted integrand {\tt sigma\_2} needed beyond
NLO. It is the variable {\tt which\_piece} that determines which of
the three functions is called. Usually, for a LO case, we only need
{\tt sigma\_0}. However, since the process $\mu\to e \nu\bar{\nu}
\gamma\gamma$ as such is not implemented in \mcmule{}, we compute it
at LO by calling the real corrections of the radiative muon decay
$\mu\to e \nu\bar{\nu} \gamma$. Thus, from a technical point of view
we call a single-subtracted integrand. The function {\tt quant},
however, is constructed such that no subtraction takes place. This is
ensured by the demand $E_{\gamma} > 10\,\MeV$. In addition, {\tt
which\_piece} determines {\tt ndim}, the dimension of the integration
(11 in our case), and the matrix element that needs to be called, {\tt
Pm2ennggAV(q1,n1,q2...q6)}. The name of the function suggests we
compute $\mu(q_1,n_1)\to e(q_2) [\nu \bar\nu] \gamma(q_5)\gamma(q_6)$
with the polarisation vector {\tt n1} of the initial lepton, and the
neutrinos are averaged over. Note that the momenta of the neutrinos
are given as arguments, even if they are redundant. This simplifies
the code a lot because it means that all matrix elements have the same
calling convention.

The interplay between the function {\tt sigma\_1(x,wgt,ndim)} and {\tt
vegas} is as usual, through an array of random numbers {\tt x} of
length {\tt ndim}. In addition the vegas weight of the event, {\tt
wgt}, is passed. The function {\tt sigma\_1} simply evaluates the
complete weight {\tt wg} of a particular event by combining {\tt wgt}
with the matrix element supplemented by symmetry, flux, and
phase-space factors.  In a first step a phase-space routine of {\tt
phase\_space} is called.  For our calculation this is the optimised
phase space {\tt
psd6\_p\_25\_26\_m50\_fks(x,p1,Mm,p2,Me,p3...p6,0.,weight)} generating
the momenta with correct masses as well as the phase-space weight {\tt
weight}. The {\tt d} in the name of the phase-space routine indicates
that we are considering a decay process (one initial state particle),
the {\tt 6} indicates the total number of momenta generated and the
meaning of {\tt fks} will be explained below. The other labels
indicate the particular tuning and partition which are irrelevant in
this case. With these momenta the observables to be computed are
evaluated with a call to {\tt quant}.  If one of them passes the cuts,
the variable {\tt cuts} is set to true.  This triggers
the computation of the matrix element and the assembly of the full
weight. In a last step, the routine {\tt bin\_it}, stored in {\tt
vegas}, is called to put the weight into the correct bins of the
various distributions.  These steps are done for all events and those
after pre-conditioning are used to obtain the final distributions.

Since, technically speaking, we are computing a subtracted matrix
element, the code also generates for each event the associated soft
event, i.e. the same event with $\xi_1\to 0$. This is realised by
having a parametrisation of the phase space, such that setting the
first entry of {\tt x} to 0 results in $\xi_1\to 0$. Such a
phase-space routine is called FKS compatible and named with the ending
{\tt fks}. It is then checked whether the subtraction condition
$\xi_1<\xi_c$, \eqref{subcond}, is satisfied. If yes, {\tt quant} is
evaluated with this new set of momenta, and if the event passes, the
soft limit of the matrix element is evaluated and the subtraction is
performed according to \eqref{subcond}.  The global variable {\tt
xiout} is required for this, since it is left out of the FKS
phase-space weight and has to be included in the integrand. In our
case, the soft event never passes the cuts, due to the requirement
{\tt q6(4) > 10.}  in {\tt quant}.

To conclude this section, we mention that the process considered here
is actually relevant to searches for lepton-flavour violating decays
mediated by a light particle $X$. Indeed, double radiative muon decay
$\mu\to e \nu\bar{\nu} \gamma\gamma$ in the region of very small
$\slashed{E}$ cannot be distinguished from $\mu\to e X$ with
$X\to\gamma\gamma$.

\begin{figure}[t]
\centering
\vspace{-6mm}\includegraphics[width=0.75\textwidth]{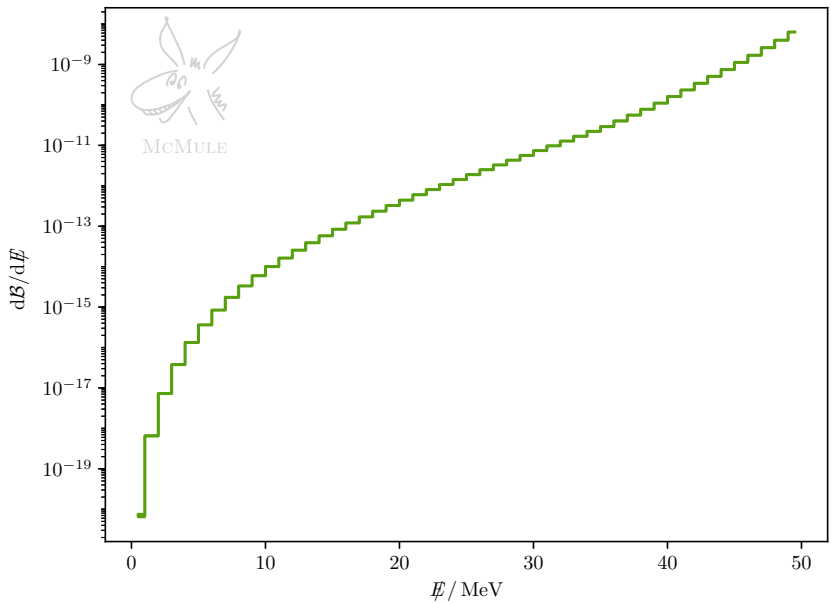}
\caption{Branching ratio at LO for the double radiative muon decay
  $\mu\to e \nu\bar{\nu} \gamma\gamma$, as a function of the missing
  energy, i.e. the energy of the neutrinos.}
\label{fig:drmd}
\end{figure}

MEG has performed a search for this decay~\cite{baldini2020search}.
In order to assess the background from double radiative muon decay, we
have computed the $\slashed{E}$ distribution, with cuts
\begin{align}
  E_{\gamma} &> 10\,\MeV, &
  |\cos\theta_{\gamma}| &< 0.35\, , &
  |\phi_{\gamma}| &> \frac{2\pi}{3}\, ,
  \label{megcut}
\end{align}
adapted to the MEG detector. Here, $\theta_{\gamma}$ and
$\phi_{\gamma}$ are the polar and azimuthal angles of the two photons.
Further, we require that the two photons can be separated in the
calorimeter. This is implemented by specifying them to be $\delta
x=20\,{\rm cm}$ apart on the detector surface which is at a radius of
$R=67.85\,{\rm cm}$ resulting in
\begin{align}
\theta_{\gamma\gamma}
> \tan^{-1}\bigg(\frac{\delta x}{R}\bigg)
\approx 16.4^\circ\,.
\end{align}
\clearpage
The results are shown in Figure~\ref{fig:drmd}, where special emphasis
has been given to the small $\slashed{E}$ region. Integrating the
differential distribution up to $\slashed{E} = 10\,\MeV$ yields a
branching ratio of $\mathcal{B}(\slashed{E}<10\,\MeV) = 1.2\times
10^{-14}$.

\section{Muon-electron scattering with \mcmule{}}
\label{sec:muone} 

Muon-electron scattering is a classic process which at low energy is
completely dominated by QED. There is renewed interest in this process
in connection with the long-standing $(3-4)\sigma$ discrepancy between
the anomalous magnetic moment or $(g-2)_\mu$ of the muon and its
Standard Model prediction. The theory calculation of $(g-2)_\mu$
suffers from uncertainties originating from non-perturbative hadronic
corrections. The largest source of this uncertainty is the HVP,
followed by the contribution due to hadronic light-by-light
scattering~\cite{Jegerlehner:2017lbd}.  A better understanding of the
hadronic contributions is therefore of utmost importance, even more so
in light of the new $(g-2)_\mu$ experiments at
Fermilab~\cite{Grange:2015fou} and J-Parc~\cite{Mibe:2011zz} that will
further increase the experimental precision achieved by the BNL E831
experiment~\cite{Bennett:2006fi}.

The HVP correction can be related to measurement data of
electron-positron annihilation using a dispersive
approach~\cite{Davier:2019can, Keshavarzi:2019abf}. The resulting
integrand is, however, highly fluctuating due to hadronic resonances
and threshold effects. This makes the corresponding analysis rather
challenging. Furthermore, a recent lattice
evaluation~\cite{Borsanyi:2020mff} of the HVP contribution to
$(g-2)_\mu$ substantially deviates from the dispersive approach.

In this context it has been  proposed to extract the HVP corrections
from the measurement of the running of the QED coupling in the
space-like region~\cite{Calame:2015fva}. Contrary to the traditional
time-like approach, the corresponding integrand is smooth and free of
resonances. Moreover, this would yield an independent determination of
the HVP contribution resulting, in turn, in a better understanding of
the theory error.

While the original proposal was based on Bhabha
scattering~\cite{Arbuzov:2004wp, Abbiendi:2005rx, Calame:2015fva}, it
was recently established that the elastic scattering of muons on
atomic electrons could, in principle, be sufficiently sensitive to
reach a competitive precision with this novel
approach~\cite{Abbiendi:2016xup}. This is the objective of the
proposed MUonE experiment~\cite{LoI}. Since in this case the effect of
the HVP to the running of the QED coupling for muon-electron
scattering ranges from $10^{-3}$ to $10^{-5}$, the differential cross
section would have to be measured at a precision of 10\,ppm.

The radiative corrections to muon-electron scattering represent one
source of systematic uncertainty that has to be carefully
studied~\cite{Banerjee:2020tdt}. To reach the target precision these
corrections have to be known at a level of 10\,ppm as well. The effect
of hadronic corrections was recently addressed
in~\cite{Fael:2018dmz,Fael:2019nsf} using two independent methods. At
leading order also $Z$-exchange has to be taken into account. The main
corrections are, however, due to QED radiation.  The minimal
requirement is expected to be the NNLO QED corrections, for which we
have to consider up to two photons in the final state
\begin{align}
  \label{muone:process}
    e^{-}(p_1)\,\mu^{-}(p_2)\rightarrow e^{-}(p_3)\,\mu^{-}(p_4)\
    \big\{\gamma(p_5)\, \gamma(p_6)\big\} \, ,
\end{align} 
matched to leading-logarithmic resummation.  In this section we report
on the progress towards this goal made through \mcmule{}.

\subsection{Running \mcmule{} for muon-electron scattering}
\label{sec:muoneRun}

The NLO QED corrections to muon-electron scattering have been known
for a long time~\cite{Bardin:1997nc,Kaiser:2010zz}. Motivated by the
MUonE experiment, they have been revisited and, together with the NLO
electroweak corrections, implemented in a fully differential Monte
Carlo code~\cite{Alacevich:2018vez}.

As a first step towards a sufficiently precise description of
muon-electron scattering within \mcmule{}, we have also implemented
the NLO QED corrections. We have compared our results with
\cite{Alacevich:2018vez} and found full agreement. \mcmule{} also
contains the dominant electronic NNLO corrections that are
proportional to $Q_\mu^2 Q_e^6$, where $Q_\mu$ and $Q_e$ denote the
charge of the muon and electron, respectively~\cite{Banerjee:2020tdt}.
Also this part of the code is fully verified after comparing the
observables defined in `Setup 2' and `Setup 4'
of~\cite{Alacevich:2018vez} with~\cite{pavia}.
Since~\cite{Alacevich:2018vez,pavia} use a photon mass as infrared
regulator and the phase-space slicing method, the agreement is a
strong cross check for a correct technical implementation. Details of
the computation and physical results will be presented in
Section~\ref{sec:muoneNNLO}. In this section we focus on a description
on how to run the code. 

There are several changes with respect to the example discussed in
Section~\ref{sec:example}. First of all, the process is different. The
generic process now is {\tt{em2em}}. For a tree-level computation we
can proceed analogous to Section~\ref{sec:example} with
{\tt{which\_piece}} set to {\tt{em2em0}}. For a NLO computation we
need to evaluate the virtual and real corrections. As shown in
\eqref{eq:nlo:4d}, using FKS this results in two terms, the subtracted
real corrections~\eqref{eq:nlo:n1} and the finite virtual
corrections~\eqref{eq:nlo:n}, i.e. the virtual corrections combined
with the infrared counterterm.  The corresponding {\tt{which\_piece}}
are {\tt{em2emR}} and {\tt{em2emF}}, respectively.

The results obtained with {\tt{em2emR}} and {\tt{em2emF}} taken
separately are $\xi_c$ dependent. This dependence has to cancel in the
sum. The $\xi_c$ parameter is set through the variable
{\tt{xinormcut1}} of Table~\ref{tab:mcmuleinput}. The latter has to be
set to a value between 0 and 1 and is related to $\xi_c$ through
\eqref{eq:xic} as {\tt{xinormcut1}}$=\xi_c/\xi_\text{max}$.  Checking
the independence of physical results on $\xi_c$ serves as a
consistency check and is an implicit check on the infrared safety of
the observable implemented in {\tt{quant}}. To do this, it helps to
disentangle {\tt{em2emF}} into {\tt{em2emV}} and {\tt em2emC},
according to \eqref{eq:nlo:n}.\footnote{This additional split is not
implemented for all processes.} The former corresponds to the pure
virtual corrections whereas the latter is the infrared counterterm,
i.e.  the integrated eikonal times the tree-level matrix element. Of
course, taken separately these terms are infrared divergent. \mcmule{}
returns the finite part, as defined in~\cite{Ulrich:2020phd,
mcmuleman}.  Only {\tt{em2emC}} depends on $\xi_c$ and this part is
typically much faster in the numerical evaluation.

\begin{figure}
\centering
\input{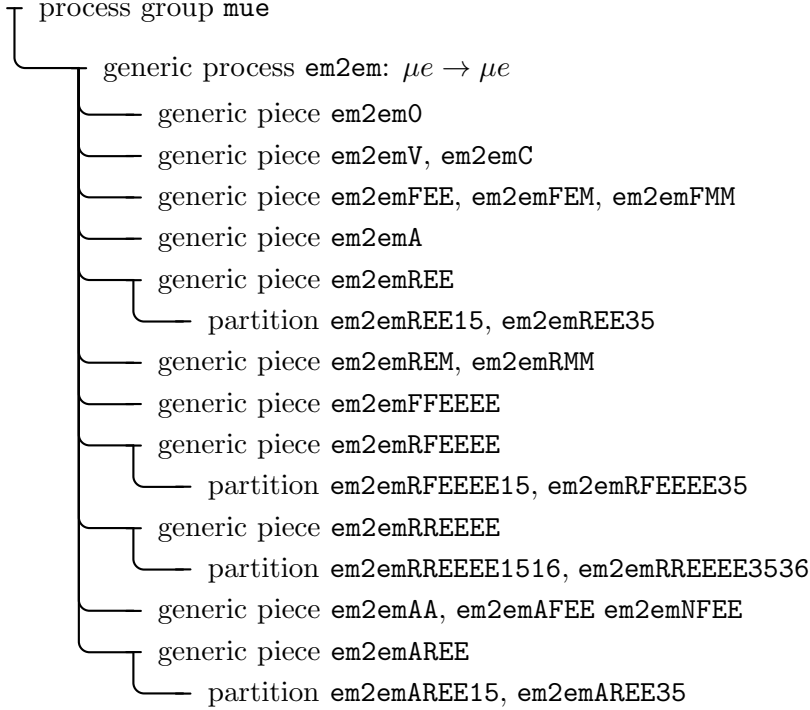}
\caption{A complete list of contributions ({\tt{which\_piece}})
  currently implemented for $\mu$-$e$ scattering. This is a subset of
  Figure~\ref{fig:processtree}.}
\label{fig:muonetree}
\end{figure}

In fact, {\tt{em2emR}} and {\tt{em2emF}} are divided up further, as
can be seen in Figure~\ref{fig:muonetree}, where a complete tree of
possible {\tt{which\_piece}} for the generic process {\tt{em2em}} is
depicted. This additional separation corresponds to a gauge-invariant
split of the NLO corrections into emission/absorption from the
electron line {\tt{EE}}, emission/absorption from the muon line
{\tt{MM}}, and the interference {\tt{EM}}. As shown
in~\cite{Alacevich:2018vez}, the {\tt{EE}} contributions are by far
dominant. In addition, these contributions suffer most from pseudo
singularities that arise from photon emission nearly collinear to the
electron. To deal with these regions of phase space in a numerically
stable way, there is one further purely technical partitioning of
{\tt{em2emREE}} into {\tt{em2emREE15}} and {\tt{em2emREE35}}. These
two partitions have a tuned phase space in $s_{15}=2 p_1\cdot p_5$ and
$s_{35}=2 p_3\cdot p_5$, respectively, to deal with initial-state and
final-state pseudo-collinear singularities.

Finally, we note that also hadronic contributions are implemented.
This is done together with the leptonic vacuum polarisation (VP) in
{\tt em2emA}. The user can then set the variables {\tt nel}, {\tt
nmu}, {\tt ntau}, and {\tt nhad} to decide which contributions to
include.  For the calculation of the HVP the Fortran library {\tt
alphaQED}\cite{Jegerlehner:2001ca, Jegerlehner:2006ju,
Jegerlehner:2011mw} is used. Specifically, we rely on the hadronic
stand-alone version {\tt hadr5n12.f}.

Choosing different random seeds, varying $\xi_c$ and having to compute
the various real and virtual parts results in quite a few jobs.  A
particularly convenient way to run \mcmule{} is using menu files. A
menu file contains a list of jobs to be computed such that the user
will only have to vary the random seed and $\xc$ as the statistical
requirements are defined globally in a config file. This is completed
by a submission script, usually called {\tt submit.sh}.  The submit
script is what will need to be launched. It will take care of the
starting of different jobs. It can be run on a normal computer or on a
Slurm cluster~\cite{Yoo:2003slurm}.

To prepare the run in this way we can use {\tt pymule}, a tool
provided together with \mcmule{}. When using {\tt pymule create}, we
are asked various questions, most of which have a default answer in
square brackets. In the end {\tt pymule} will create a directory,
where all results will be stored.  In addition {\tt pymule} also
provides tools to analyse the results, such as combining runs with
different random seeds and different choices of $\xi_c$. A more
detailed description of {\tt{pymule}} can be found in the online
documentation~\cite{mcmuleman, Ulrich:2020phd}.

Moving from NLO to NNLO increases the number of partial results
further. Now we have to run with {\tt{which\_piece}} set to
{\tt{em2emRREEEE}} (double-real corrections), {\tt{em2emRFEEEE}} (real
virtual corrections), and {\tt{em2emFFEEEE}} (double-virtual
corrections). As discussed above, the additional ending {\tt{EEEE}}
indicates that only electronic corrections are included.  As for the
real corrections, also the real-virtual and the double-real
contributions are computed with a partition to disentangle
initial-state and final-state pseudo-collinear singularities. 

Also at NNLO the correction due to hadronic and leptonic VP is
included. These contributions are split up according to the
classification of~\cite{Fael:2019nsf}. The diagrams where the VP
factorises are implemented in {\tt em2emAA}, {\tt em2emAR}, and {\tt
em2emAF}. The former takes into account diagrams with one or two
insertions of the VP into the tree level diagram. The latter two
implement QED NLO corrections combined with one insertion. The
remaining non-factorisable vertex correction can be computed via {\tt
em2emNF}. This relies on the results of~\cite{Fael:2018dmz} which uses
the hyperspherical integration method to calculate the hadronic
corrections to muon-electron scattering~\cite{Laporta:1994mb}.

\begin{figure}
\centering
\includegraphics[width=0.75\textwidth]{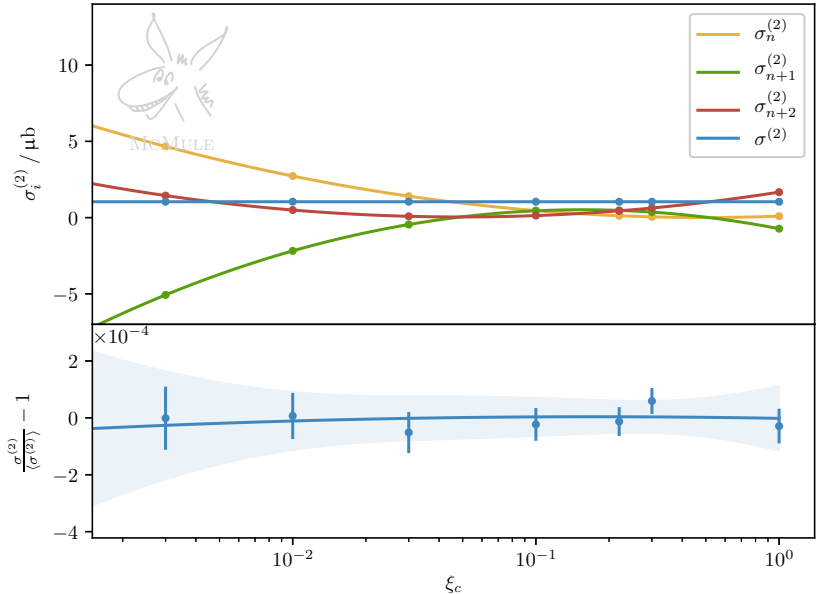}
\caption{The (in)dependence of the pure NNLO contribution
  $\sigma^{(2)}$ to the total cross section on the arbitrary cut
  parameter $\xi_c$. The error band represents the $1\sigma$
  confidence level of the fit.}
\label{fig:xicut}
\end{figure}

As listed in Table~\ref{tab:mcmuleinput}, running at NNLO there are
two $\xi_c$ variables to be set in the input. However, to obtain
$\xi_c$ independent physical results it is imperative that they are
set equal, {\tt{xinormcut1}}={\tt{xinormcut2}}. The reason \mcmule{}
works with two variables is that for computations with massless
fermions, {\tt{xinormcut2}} corresponds to the unphysical cut variable
related to the collinear subtraction, often denoted by
$\delta_{\text{cut}}$. Also, an independent {\tt{xinormcut2}} can be
used for internal checks.

An example of a typical check of the $\xi_c$ (in)dependence is shown
in Figure~\ref{fig:xicut}, where the $n$-particle (orange),
$(n+1)$-particle (green), and $(n+2)$-particle (red) contributions are
shown separately for the total cross section according to `Setup 4'
of~\cite{Alacevich:2018vez}.\footnote{In fact, this was one of the
numbers compared with~\cite{pavia}.} These are just the parts given in
\eqref{eq:nnlo:4d}. In the sum (blue) the $\xi_c$ dependence cancels.
This can be seen particularly well in the bottom panel, where the
results of seven separate choices of $\xi_c$ are shown, together with
a 1$\sigma$ band of a fit to the $\xi_c$ dependence.

Once more we stress that this cancellation is exact. Thus, in
principle any choice is allowed. However, for very small choices of
$\xi_c$ there are large numerical cancellations. In the case of
production runs, it is thus advisable to pick a value of $\xi_c$ where
the separate contributions have roughly the same magnitude as the
final result. From experience, a choice around $\xi_c \sim 0.1$ is a
good starting point.

\subsection{The dominant NNLO corrections}
\label{sec:muoneNNLO}

We now turn to the technical details of the calculation as well as the
phenomenological discussion of the results. As previously mentioned,
at NNLO we restrict ourselves to the gauge-invariant subgroup that
only contains electronic corrections, i.e. contributions proportional
to $Q_\mu^2 Q_e^6$. These corrections are expected to be dominant
compared to the other contributions at this perturbative order as a
consequence of enhanced collinear logarithms. To be consistent, at
NNLO we therefore also only include VP with electrons inside the loop.

Furthermore, we assume the electron to be unpolarised in
correspondence with the atomic electrons of the MUonE experiment. Our
results are therefore independent of the muon polarisation due to
parity invariance of QED. Additionally, the considered gauge-invariant
subset is also independent of whether the muon beam consists of
$\mu^+$ or $\mu^-$. This does not hold, however, for the full set of
NLO QED corrections that is included here. This includes the muon and
tau VP.

The double-virtual diagrams were calculated with the full electron
mass dependence using the analytic expressions for the heavy quark
form factors of~\cite{Bernreuther:2004ih}. Furthermore, the genuine
two-loop corrections to the photon self-energy were taken
from~\cite{Djouadi:1993ss}. The diagrams for the real-virtual and
double-real contributions were generated using
QGraf~\cite{Nogueira:1991ex} and calculated with
Package-X~\cite{Patel:2015tea}. An independent calculation was
performed using FORM~\cite{Ruijl:2017dtg}. Complicated scalar
triangle- and box-functions were then evaluated with the {\tt COLLIER}
library~\cite{Denner:2016kdg}. Additionally, {\tt COLLIER} was used to
perform a numerical stable tensor reduction in problematic regions of
the phase space.

With the momenta of the particles labelled as in \eqref{muone:process}
we define the invariants $t_e=(p_1-p_3)^2$ and $t_\mu=(p_2-p_4)^2$.
In the case of purely virtual corrections we have $t_e=t_\mu$.  The
energy of the outgoing electron and muon are denoted by $E_e$ and
$E_\mu$, respectively. Additionally, we use $\theta_e$ and
$\theta_\mu$ as the corresponding scattering angles relative to the
beam axis. We further assume a muon beam of energy $E=150\ \GeV$,
consistent with the M2 beam line at CERN North Area\cite{LoI}.

The total cross section is ill-defined due to the behaviour
$d\sigma/dt \sim t^{-2}$ with $t_{\rm min}\leq t\leq 0$. We therefore
have to apply a cut on the maximal value of $t$ or equivalently on the
minimal energy of the outgoing electron. In all of the results below
we have chosen $E_e > 1\,\GeV$. To model the geometry of the detector
we require in addition that $\theta_\mu > 0.3\ {\rm mrad}$. 

Following~\cite{Banerjee:2020tdt}, the outgoing electron and muon
angles are in the absence of photons related through the elasticity
condition
\begin{align}
    \tan{\theta_\mu^{\rm el}} = \frac{2\tan{\theta_e}}{(1+\gamma^2 \tan^2{\theta_e})(1+g_\mu^\ast)-2}
    \,,
\end{align}
where
\begin{align}
    g_\mu^\ast = \frac{E m +M^2}{E m + m^2}, \quad\quad \gamma = \frac{E+m}{\sqrt{s}}
\end{align}
with $s$ the centre-of-mass energy. This allows to restrict radiation
with the elasticity cut
\begin{align}
    0.9 < \frac{\theta_\mu}{\theta_\mu^{\rm el}} < 1.1\, .
\end{align}
In the following we present results with and without this additional
cut, in order to analyse its impact on the radiative corrections. A
similar effect can be expected as in the case of the acoplanarity cut
of~\cite{Alacevich:2018vez}, where the NLO corrections flatten out
significantly. 

\begin{table}
\centering
 \begin{tabular}{c|r r|| r r} 
 & \multicolumn{2}{c||}{$\sigma/\rm \upmu b$} & \multicolumn{2}{c}{$\delta K^{(i)}/\%$} \\
  & \multicolumn{1}{c}{\tt S1} & \multicolumn{1}{c||}{\tt S2} & \multicolumn{1}{c}{\tt S1} & \multicolumn{1}{c}{\tt S2} \\ [0.5ex] 
 \hline
 $\sigma^{(0)}$ & \tt 121.4229  &\tt 121.4229& & \\ 
 \hline
 $\sigma^{(1)}$ & \tt   0.5440  &\tt  -4.0773& \tt 0.4480&\tt -3.3580\\
 \hline
 $\sigma^{(2)}$ & \tt  -0.0058  &\tt  +0.0093& \tt-0.0048&\tt  0.0079 \\
 \hline\hline
 $\sigma_{2}$   & \tt 121.9611  &\tt 117.3549& & \\
\end{tabular}

\caption{Results for the integrated cross section for {\tt S1}
(without elasticity cut) and {\tt S2} (with elasticity cut) at LO,
NLO, and NNLO. All digits given are significant compared to the error
of the numerical integration.}
\label{tab:muone-integrated}
\end{table}

\begin{figure}
\centering
\subfloat[{\tt S1}]{
\includegraphics[width=0.75\textwidth]{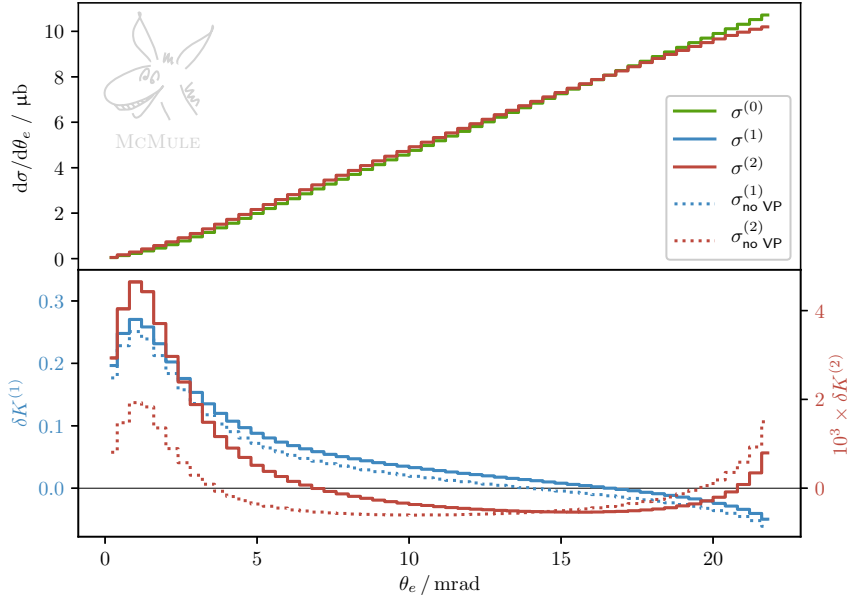}
\label{theta-e-no-cut}
}\\
\subfloat[{\tt S2}]{
\includegraphics[width=.75\textwidth]{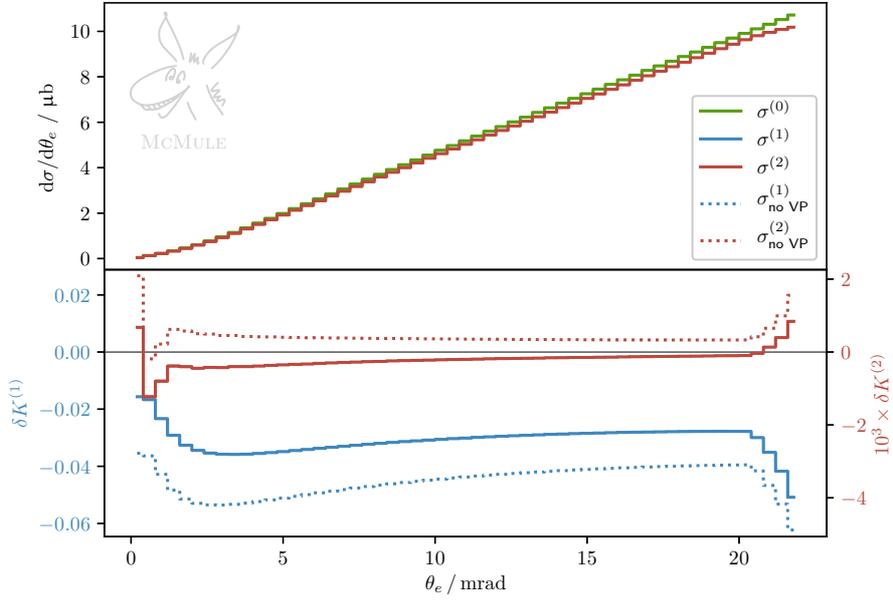}
\label{theta-e-yes-cut}
}
\caption{The differential cross section w.r.t. $\theta_e$ at LO
(green) and NNLO (red) for scenarios {\tt S1} (without elasticity cut)
and {\tt S2} (with elasticity cut). The NLO and NNLO $K$ factors are
shown in blue and red, respectively.}
\label{fig:muone-differential-theta}
\end{figure}

\begin{figure}
\centering
\subfloat[{\tt S1}]{
\includegraphics[width=.75\textwidth]{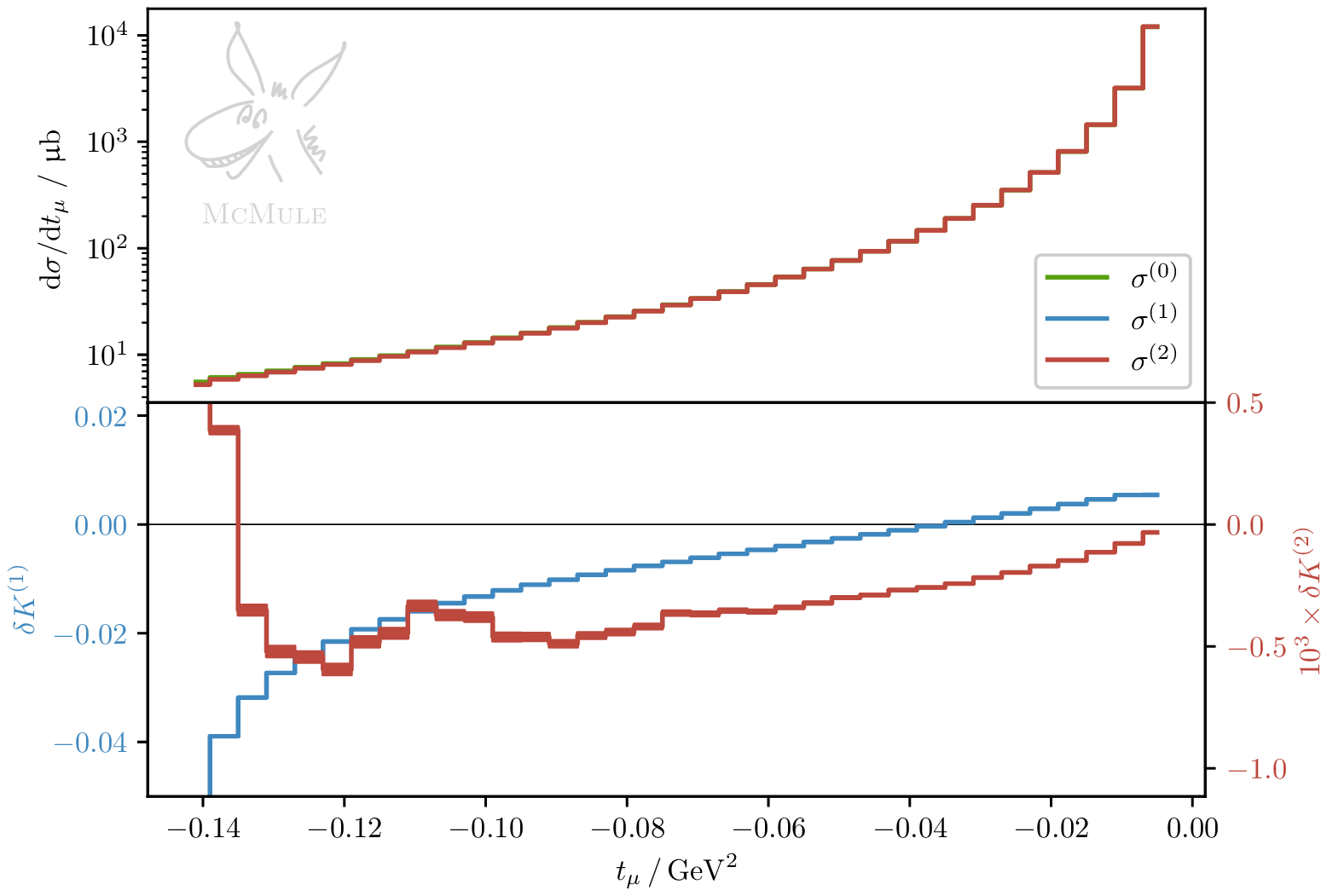}
\label{tmm-no-cut}
}\\
\subfloat[{\tt S2}]{
\includegraphics[width=.75\textwidth]{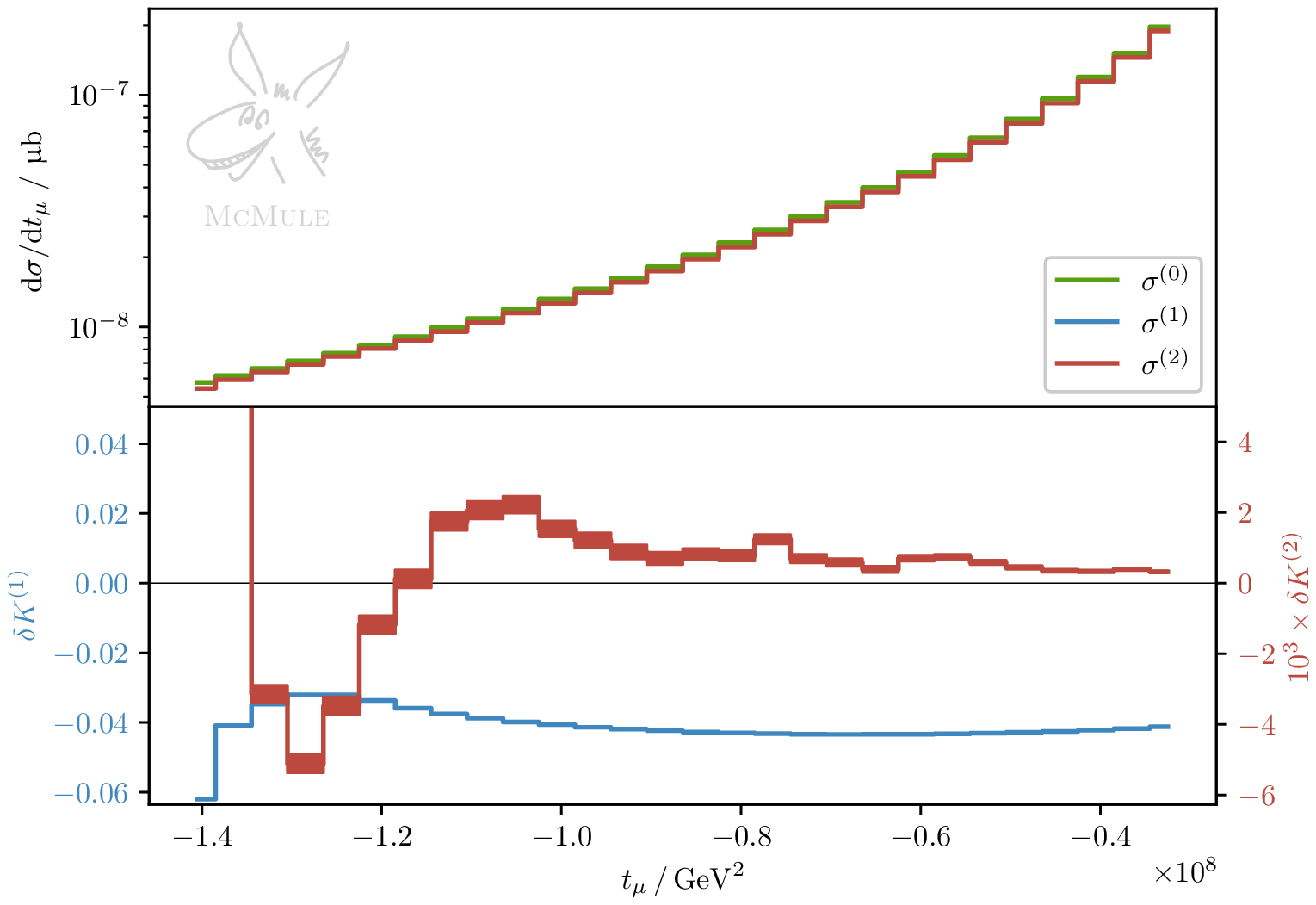}
\label{tmm-yes-cut}
}
\caption{The differential cross section w.r.t. $t_\mu$ at LO (green)
and NNLO (red) for scenarios {\tt S1} (without elasticity cut) and
{\tt S2} (with elasticity cut). The NLO and NNLO $K$ factors are shown
in blue and red, respectively.}
\label{fig:muone-differential-t}
\end{figure}

In summary, we consider the two scenarios
\begin{itemize}
    \item {\tt S1}: $E_e > 1\,\GeV$, $\theta_\mu > 0.3\ {\rm mrad}$,
    \item {\tt S2}: $E_e > 1\,\GeV$, $\theta_\mu > 0.3\ {\rm mrad}$, 
          $0.9 < \theta_\mu / \theta_\mu^{\rm el} < 1.1$.
\end{itemize}
The order-by-order contributions, $\sigma^{(i)}$, to the integrated
cross section, $\sigma_2=\sigma^{(0)} + \sigma^{(1)} + \sigma^{(2)}$,
are presented in Table~\ref{tab:muone-integrated}.\footnote{For this
paper, $\sigma^{(2)}$ only denotes the dominant NNLO contribution as
defined in the various sections.} It also shows the corresponding $K$
factors defined as
\begin{align}
    K^{(i)}=1+\delta K^{(i)} =
    \frac{\sigma_{i}}{\sigma_{i-1}}
\end{align}
Figure~\ref{fig:muone-differential-theta} and
Figure~\ref{fig:muone-differential-t} then show differential results
that are of interest to the MUonE experiment. In particular, we
present distributions with respect to $\theta_e$ and $t_\mu$. The
differential cross section at LO as well as at NNLO are displayed in
the upper panels. In addition, the lower panels show the differential
$K$ factors
\begin{align}
   \delta K^{(i)}=
   \frac{d\sigma^{(i)}/dx}{d\sigma_{i-1}/dx}
\end{align}
with $x\in\{\theta_e,t_{\mu}\}$. In dotted lines, the $K$ factors
without the inclusion of the VP are shown.

We first remark that the numerical error for the distribution
$d\sigma/d\theta_e$ (Figure~\ref{fig:muone-differential-theta}) is
much smaller than for $d\sigma/d t_\mu$
(Figure~\ref{fig:muone-differential-t}). This is due to the fact that
the cross section in the latter case is practically zero in most parts
of the kinematically allowed region. As exemplified in
Section~\ref{sec:example}, the statistics for $d\sigma/d t_\mu$ could
be drastically improved using tailored runs. Nevertheless, the
discontinuities of Figure~\ref{fig:muone-differential-t} indicate that
the Monte Carlo error for individual bins provided by \mcmule{} might
be underestimated.

Furthermore, sizeable NLO and NNLO corrections of up to $30\%$ and
$0.5\%$, respectively, can be observed. Naively, one could therefore
conclude that the target precision of 10\,ppm of MUonE is far out of
reach. First of all, however, it has to be noted that the enhancement
of the corrections at the end points of the distributions are due to
soft photon emission. For a reliable description in this region, the
logarithms need to be resummed. The leading logarithms can be resummed
with a parton shower.  Moreover as detailed
in~\cite{Banerjee:2020tdt}, also the calculation of the
next-to-leading logarithms to all orders might be feasible. Secondly,
the elasticity cut has the important effect of significantly reducing
the variation in the $K$ factors. Since the MUonE experiment proposes
to measure ratios of cross sections of different kinematic regions to
cancel systematic uncertainties as opposed to absolute values, the
flatness of the corrections is highly
advantageous~\cite{Banerjee:2020tdt, Masiero:2020vxk}.


\section{Lepton-proton scattering with \mcmule{}}
\label{sec:muse} 

There is a long history in the study of elastic electron-proton and
muon-proton scattering and the computation of radiative corrections to
these processes started in the sixties~\cite{Mo:1968cg}. If
lepton-mass effects are taken into account appropriately, the two
processes are the same from a computational point of view.

Typically, the corrections to lepton-proton scattering at
NLO~\cite{Maximon_2000, Akushevich:2015toa} are split into different
gauge-independent parts: corrections from the lepton line, corrections
due to VP effects, corrections from the proton line and the so-called
two-photon exchange corrections~\cite{Arrington:2011dn,
  Afanasev:2017gsk,  Tomalak:2015hva, Tomalak:2015aoa,
  Tomalak:2017shs}. The latter contain contributions
from inelastic intermediate states. This makes it difficult to obtain
solid predictions from first principles. Looking at the charge
asymmetry, i.e. the difference between $\ell^+\, p$ and $\ell^-\, p$
scattering, is a useful tool to gain more information on two-photon
exchange~\cite{Koshchii:2017dzr}.

Going beyond NLO, the situation becomes considerably more complicated.
To mention just two complications, the VP effects cannot be factorised
any longer~\cite{Fael:2019nsf} and apart from the emission of photons
also the emission of an additional $\ell^+ \ell^-$ pair potentially
needs to be considered.

Due to the small mass of the electron, corrections from the lepton
line are typically dominant for $e p\to e p$. Hence, they received
particular attention. Effects beyond the the soft approximation for
the radiation from the electron were
considered~\cite{Weissbach:2008nx}, as well as resummation of leading
logarithmic effects~\cite{Arbuzov:2015vba}. Recently a calculation
including the corresponding  NNLO effects was presented in
\cite{Bucoveanu:2018soy}. Resummation was also studied in an effective
theory approach~\cite{Hill:2016gdf}. 

For these corrections, the only difference between lepton-proton
scattering and the results presented in the previous section is due to
the fact that the proton is not pointlike. This can be accounted for
by parametrising the photon-proton interaction through form factors.
Of course, electron-proton scattering has been used to determine these
form factors and, in particular, their behaviour for small momentum
transfer squared. This allows for an extraction of the proton radius,
see e.g.~\cite{Lee:2015jqa}.  However, for the results presented in
this section we will simply use the standard dipole form factors, as
given in Appendix~\ref{sec:appendix}.

Thus, in this section we show NNLO results for unpolarised elastic
lepton-proton scattering
\begin{align}
  \label{lp:process}
  \ell(p_1)\, p(p_2) \to \ell(p_3)\, p(p_4)\
    \big\{\gamma(p_5)\, \gamma(p_6)\big\} 
\end{align}
in the approximation that the lepton interacts with the proton through
the exchange of a single photon with the standard dipole form factor.
All lepton mass effects as well as leptonic and hadronic VP effects
are taken into account.  On the other hand, two (or more) photon
exchange as well as radiation from the proton is neglected. We also
make the assumption that there are no additional lepton pairs in the
final state.

\subsection{NNLO effects in elastic electron-proton scattering}
\label{subsec:p2}

As a first example, we consider $e^-\, p \to e^-\, p$ in a setting
with kinematics adapted to the P2 experiment~\cite{Becker:2018ggl}. An
incoming electron of energy $E = 155\,\MeV$ is scattering off a proton
initially at rest.  We consider scattering angles in the range
$25^\circ< \theta_e< 45^\circ$.  Following~\cite{Bucoveanu:2018soy},
we also apply a cut on the energy of the outgoing electron and require
$E_e > 45\,\MeV$.

Starting with the total cross section (subject to the cuts above) we
list the results in Table~\ref{tab:p2}. Apart from listing the full
NLO and NNLO corrections, $\sigma^{(1)}$ and $\sigma^{(2)}$, we also
give separately the VP contribution (leptonic and hadronic) to the NLO
and NNLO corrections. While the NLO corrections are rather large
(about 5\% with the VP contributing about 1\%) the NNLO corrections
are below 0.1\%. 

\begin{table}
\centering
 \begin{tabular}{c|r||r} 
 & \multicolumn{1}{c||}{$\sigma/\rm \upmu b$} & \multicolumn{1}{c}{$\delta K^{(i)}/\%$} \\
 \hline
 $\sigma^{(0)}$             & \tt 34.5392 & \\ \hline
 $\sigma^{(1)}$             & \tt  1.7763  & \tt 5.1430\\
 $\sigma^{(1)}_\text{VP}$   & \tt  0.4663  & \tt 1.3501 \\\hline
 $\sigma^{(2)}$             & \tt -0.0237  & \tt -0.0653 \\
 $\sigma^{(2)}_ \text{VP} $ & \tt  0.0132  & \tt  0.0364 \\
 \hline\hline
 $\sigma_{2} $ & \tt 36.2919  &\\
\end{tabular}
\caption{\label{tab:p2} Results for the integrated cross section for the P2 setting at LO, NLO, and NNLO.}
\end{table}

\begin{figure}
\centering
\includegraphics[width=0.75\textwidth]{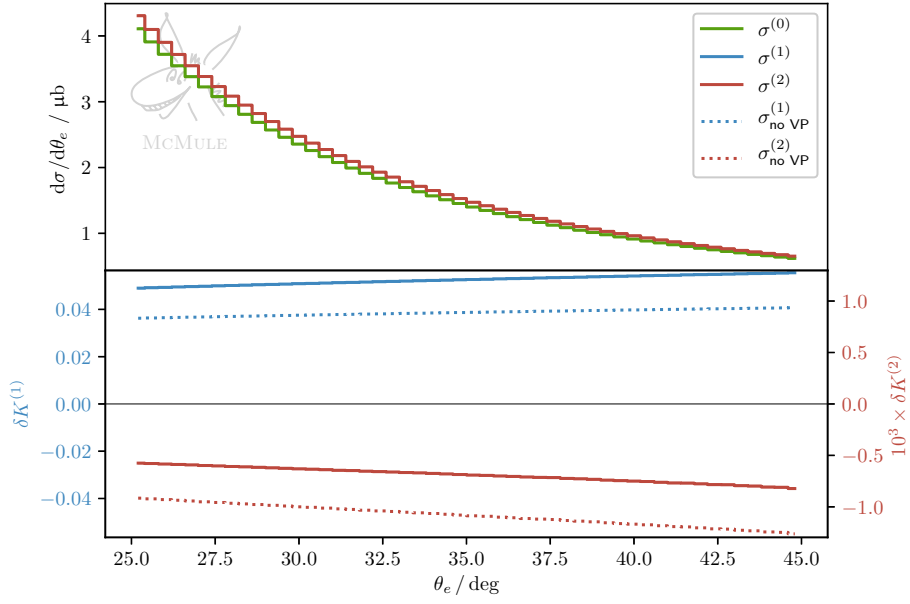}
\caption{Differential cross section $d\sigma/d\theta_e$ for a P2
  setting at LO (green) and NNLO (red) with $K$ factors. Solid
  (dotted) $K$ factors are with (without) the inclusion of VP
  contributions. }
\label{fig:mesathe}
\end{figure}

The first differential observable we consider is $d\sigma/d\theta_e$.
In the top panel of Figure~\ref{fig:mesathe} we show the LO (green)
and NNLO (red) differential cross section. The latter includes VP
contributions. In order to assess the effect of higher-order
corrections we show the $K$ factors in the bottom panel. The solid
(dotted) lines refer to the corrections with (without) VP
contributions.  Since there are no large logarithms for this
observable, the size of the corrections is in agreement with the
expectation due to the counting of powers of $\alpha$ for all values
of $\theta_e$. Consequently, the missing N$^3$LO contributions due to
emission from the electron are expected to be $\mathcal{O}(10^{-6})$
and, hence, negligible. Emission from the proton and two-photon
exchange contributions, however, will need to be properly taken into
account.

Results similar to those shown in Figure~\ref{fig:mesathe} have been
presented in~\cite{Bucoveanu:2018soy}, not including VP contributions.
Our NLO results (without VP) agree with these results. However, we
disagree substantially with the NNLO corrections
of~\cite{Bucoveanu:2018soy}, even if we adapt to their calculation and
include the electron loop in the two-loop vertex diagram. In fact, our
NNLO corrections are negative, whereas those presented
in~\cite{Bucoveanu:2018soy} are positive.  With respect to the results
presented in Section~\ref{sec:muoneNNLO} that have been verified
independently by~\cite{pavia}, the only new ingredients are the matrix
elements. They have been compared pointwise
with~\cite{Bucoveanu:2018soy} and agree.

\begin{figure}
\centering
\includegraphics[width=0.75\textwidth]{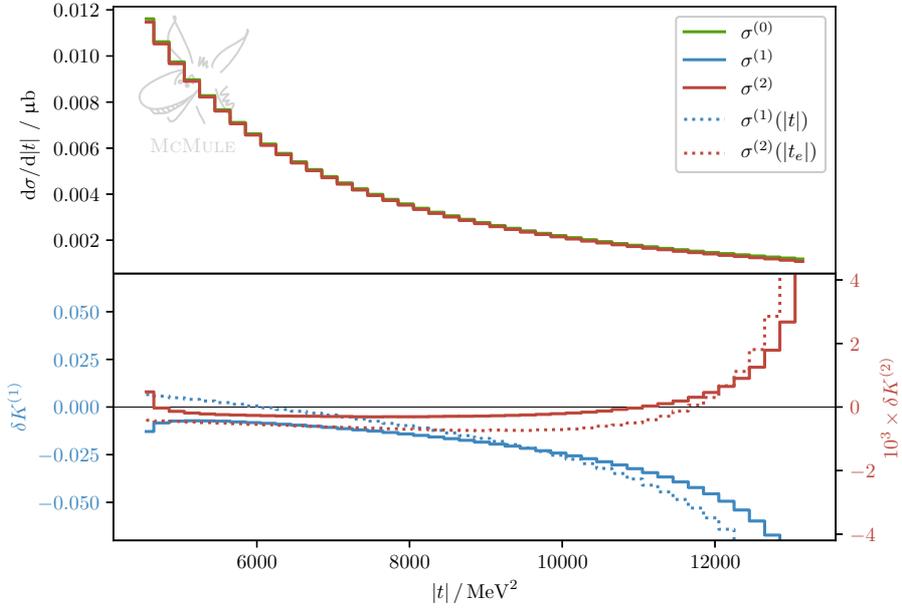}
\caption{Differential cross section $d\sigma/d|t|$ for a P2 setting at
  LO (green) and NNLO (red) with $K$ factors. Solid (dotted) $K$
  factors are determined from proton (electron) kinematics. }
\label{fig:mesaqsq}
\end{figure}

As a second example we show $d\sigma/d|t|$ in
Figure~\ref{fig:mesaqsq}. The difference between the LO result (green)
and NNLO result (red) is barely visible in the top panel. The size of
the higher-order correction can be read off from the lower panel.
While at LO, $t = (p_2-p_4)^2$ determined from the proton kinematics
is the same as $t_e\equiv{(p_1-p_3)^2}$ determined from electron
kinematics, these two quantities start to differ at NLO. In our
approximation the 'true' $Q^2$ that enters the form factors is
$Q^2=-t$. To illustrate the difference, we show the $K$ factors with
$|t|$ as well as the electronic $|t_e|$. The size of the corrections
differs by about $20\%$ between the two observables.

Generally speaking, the corrections are well under control for most
values of $Q^2=|t|$, but increase towards the endpoint as in
Figure~\ref{fig:muone-differential-t}.  Indeed, NLO (NNLO) correction
up to $10$\% ($0.5$\%) are found in the tail of the distribution, and
to obtain a very precise theoretical prediction in this region, large
logarithms would have to be resummed.

\subsection{NNLO effects in elastic muon-proton scattering}
\label{subsec:muse}

Elastic muon-proton scattering $\mu p\to \mu p$ can be used to obtain
an independent extraction of the proton radius and shed light on
possible differences between muons and electrons. In fact,
MUSE~\cite{Gilman:2017hdr} will measure simultaneously $\ell^\pm$-$p$
scattering with $\ell\in\{e,\mu\}$.  Since we are neglecting
two-photon exchange, there is no difference between $\ell^+$ and
$\ell^-$ and the only difference to the process of
Section~\ref{subsec:p2} is the mass of the lepton. As we will see
below, the larger mass of the muon typically results in smaller
corrections.

For the purpose of illustration we consider an incoming muon of
momentum $|\vec{p}_1|=210\,\MeV$ scattering off a proton at rest. For
the scattering angle range we use $20^\circ<\theta_\mu<100^\circ$, as
appropriate for MUSE. We include the same contributions as in
Section~\ref{subsec:p2}. Again we start with the total cross section
(subject to the cuts above) and present our results in
Table~\ref{tab:muse}. In this case, the NLO (NNLO) corrections are
just over $10^{-2}$ $(10^{-4})$ and are actually dominated by the VP
contributions.

\begin{table}
\centering
 \begin{tabular}{c|r||r} 
 & \multicolumn{1}{c||}{$\sigma/\rm \upmu b$} & \multicolumn{1}{c}{$\delta K^{(i)}/\%$} \\
 \hline
 $\sigma^{(0)}$             & \tt 49.6677 & \\ \hline
 $\sigma^{(1)}$             & \tt  0.6541  &\tt  1.3170\\
 $\sigma^{(1)}_\text{VP}$   & \tt  0.7172  &\tt  1.4440 \\\hline
 $\sigma^{(2)}$             & \tt  0.0075  &\tt  0.0150 \\
 $\sigma^{(2)}_ \text{VP} $ & \tt  0.0076  &\tt  0.0151 \\
 \hline\hline
 $\sigma_{2} $ & \tt 50.3294  &\\
\end{tabular}
\caption{\label{tab:muse} Results for the integrated cross section for the MUSE setting at LO, NLO, and NNLO.}
\end{table}

\begin{figure}
\centering
\includegraphics[width=0.75\textwidth]{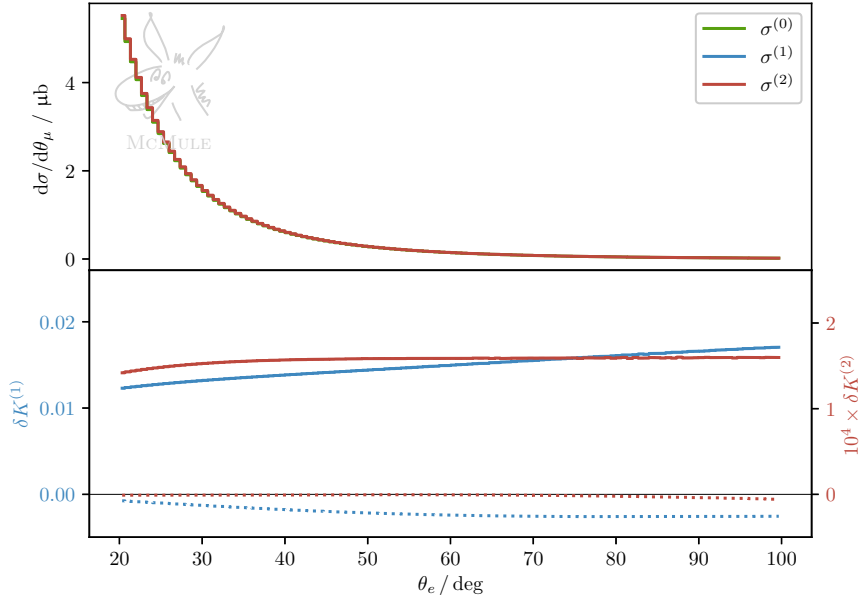}
\caption{Differential cross section $d\sigma/d\theta_\mu$ for MUSE
  with incoming muons of momentum 210\,\MeV, at LO (green) and NNLO
  (red) with $K$ factors. Solid (dotted) $K$ factors are with
  (without) the inclusion of VP contributions. }
\label{fig:musethmu}
\end{figure}

The results for the differential cross section $d\sigma/d\theta_\mu$
are depicted in Figure~\ref{fig:musethmu}. Again we show the
$K$~factor with (solid) and without (dotted) VP contributions. This
shows the dominance of the VP effects which themselves are entirely
driven by the contribution of the electron. The corrections are
roughly a factor 4 smaller than for electron-proton scattering shown
in Figure~\ref{fig:mesathe}.  Accordingly, we expect N$^3$LO
corrections from the emission of the muon to contribute well below
$\mathcal{O}(10^{-6})$ to $d\sigma/d\theta_\mu$. This is encouraging
in particular if these effects are seen as background to measure and
study two-photon contributions.

\begin{figure}
\centering
\includegraphics[width=0.75\textwidth]{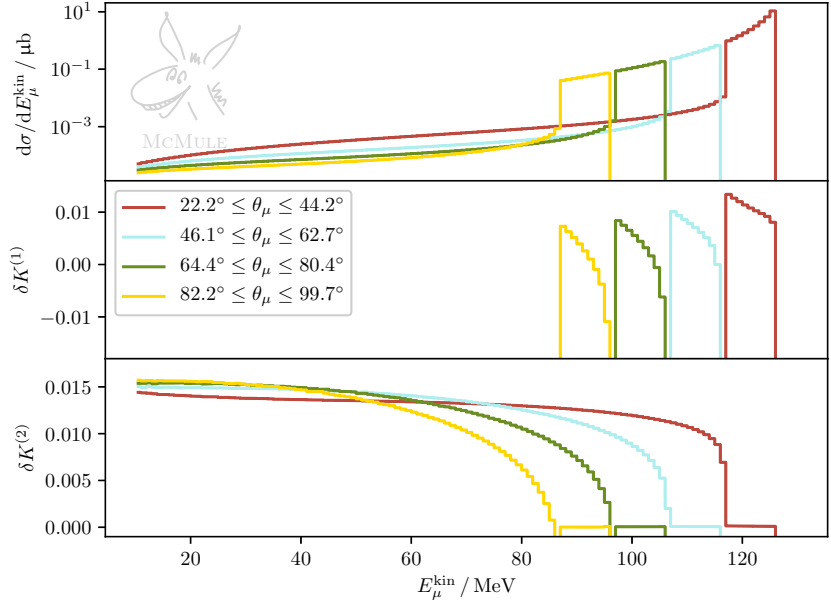}
\caption{Differential cross section $d\sigma/dE^{\text{kin}}_\mu$ for
  MUSE with incoming muons of momentum 210\,\MeV, at LO (green) and
  NNLO (red) with $K$ factors. Results are shown separately for
  different bands of the scattering angle $\theta_\mu$. }
\label{fig:museband}
\end{figure}

As a second differential observable we consider
$d\sigma/d{E^{\text{kin}}_\mu}$, where the kinetic energy of the muon
is defined as $E^{\text{kin}}_\mu \equiv E_\mu-m_\mu$. At LO there is
a one-to-one relation between the scattering angle $\theta_\mu$ and
$E^{\text{kin}}_\mu$. Beyond LO, for a given $\theta_\mu$ there will
be events with smaller $E^{\text{kin}}_\mu$ due to additional
radiation. In order to illustrate this, we define four $\theta_\mu$
bands as follows:
\begin{align}
  \label{thmuband}
  \begin{aligned}
     &\mbox{band 1}:& 22.206^\circ < \theta_\mu < 44.169^\circ &
    &  126\,\MeV &> E^{\text{kin}}_\mu|_\text{LO} > 117\,\MeV \\
    &\mbox{band 2}:& 46.148^\circ < \theta_\mu < 62.678^\circ &
    &  116\,\MeV &> E^{\text{kin}}_\mu|_\text{LO} > 107\,\MeV \\
    &\mbox{band 3}:& 64.443^\circ < \theta_\mu < 80.402^\circ &
    &  106\,\MeV &> E^{\text{kin}}_\mu|_\text{LO} > \phantom{1}97\,\MeV \\
    &\mbox{band 4}:& 82.222^\circ < \theta_\mu < 99.663^\circ &
    &  96\,\MeV &> E^{\text{kin}}_\mu|_\text{LO} > \phantom{1}87\,\MeV 
  \end{aligned}
\end{align}
The corresponding values for $E^{\text{kin}}_\mu$ at LO are also
indicated. At LO, all events of a given band will fall into this range
of $E^{\text{kin}}_\mu$. This can be seen in the top panel of
Figure~\ref{fig:museband}, where $d\sigma/d{E^{\text{kin}}_\mu}$ at
NNLO is shown in red (band 1), azure (band 2), green (band 3), and
yellow (band 4). Outside the LO $E^{\text{kin}}_\mu$ range, the cross
section falls sharply and is only non-zero due to radiative events.
The middle panel shows the NLO $K$ factor. Since $K^{(1)}$ is formally
infinity outside the LO $E^{\text{kin}}_\mu$ range, this factor is
only shown in the region where the LO cross section does not vanish.
Finally, in the lowest panel we show the NNLO $K$ factor. Within the
LO $E^{\text{kin}}_\mu$ range, these corrections are small in
accordance with the $\alpha^2$ suppression. Outside the LO
$E^{\text{kin}}_\mu$ range, however, the NNLO corrections are quite
large, up to 1.5\%. This is not very surprising, since in this
kinematic regime the NNLO terms are in fact only a NLO description of
the observable.

\section{Future developments of \mcmule{}}
\label{sec:future}

Once a mule has made up its mind, it is difficult to stop it. Hence,
there will be continuous further developments and extensions of the
code.

Roughly speaking, further developments can be divided into two
classes. First, new processes or more complete descriptions of already
implemented ones will be added. Second, there will be technical
advances that improve the performance and precision, and potentially
enable the implementation of more complicated processes.

With respect to the first class, work is ongoing to implement
M\o{}ller scattering and $e^+\, e^-\to \gamma\, \gamma$ at NNLO. As an
example for improvements on already implemented scattering processes
we mention the inclusion of polarisation effects. So far, only the
descriptions of decays are available for polarised initial states.
Many of the processes mentioned in the introduction are, however,
related to measurements of asymmetries that typically require
polarised initial states. Such observables also often require the
inclusion of electroweak corrections, as they lead to parity-violating
effects. In addition, a full NNLO description of muon-electron
scattering is envisaged. In order to go beyond the approximation of
electronic corrections the full two-loop matrix element is required.
The corresponding integrals in the limit of massless electrons are
known~\cite{Mastrolia:2017pfy, DiVita:2018nnh} and the amplitude is
being computed~\cite{Ronca:2019kcw}. In addition, also the one-loop
matrix element for $e\mu\to e\mu\gamma$ is required.  The
implementation of one-loop amplitudes for NNLO calculations requires
particular care, since they are to be integrated over singular corners
of the phase space. This results in two requirements. First, they have
to be implemented with extreme numerical stability. Second, the
numerical evaluation has to be reasonably fast.

To address these issues, in the long term it is probably advisable to
link \mcmule{} to a dedicated code that evaluates higher-order
amplitudes. There are several one-loop tools that specialise in this
(for example~\cite{Cullen:2014yla, Alwall_2014, Actis:2016mpe,
  Buccioni:2017yxi}). While so far all attempts regarding automated
computations of one- and two-loop amplitudes were dedicated to
high-energy processes, it should be possible to adapt these tools to
QED computations with massive fermions.
OpenLoops~\cite{Buccioni:2019sur} is one such tool that in the past
has been relied upon for real virtual corrections. We also plan to set
up an interface to OpenLoops to facilitate their numerically stable
calculation. Of course, a major hurdle on the path towards using an
external tool for all amplitudes is that the tool would have to be
extended to two-loop calculations.  While first steps have been made
in this direction~\cite{Pozzorini:2020hkx}, we anticipate that
two-loop amplitudes will have to be implemented directly in \mcmule{}
for the time being.

Also related to the numerical stability of the integration is the
treatment of pseudo singularities related to near collinear emission
of photons.  This is dealt with by splitting up the phase space such
that only a small number (ideally one) of pseudo singularities is
possible in each partition. Then the phase space is tuned such that
there is a simple one-to-one match between the dangerous regions of
phase space and an integration variable. Such a phase-space
parametrisation typically results in a stable and reliable numerical
evaluation of the integrals. As a possible further development, there
is the option to subtract the pseudo-collinear singularity and add
back a partially integrated counterterm~\cite{Dittmaier:1999mb}.
However, since the logarithms arising from these phase-space region
are physical, it is important to have a very flexible and exclusive
treatment of the final-state particles.

Since FKS$^\ell$ works at all orders in perturbation theory, it is
only the lack of the matrix elements that prevents us from going
beyond NNLO. One example, where a N$^3$LO calculation might be
feasible in the near future concerns the dominant electronic
contribution to muon-electron scattering. As a more futuristic
development we mention the idea to possibly compute the finite
(eikonal-subtracted and ultraviolet-renormalised) matrix elements
$\fM{n+i}{\ell-i}$ that are the ingredients of FKS$^\ell$ directly
numerically.

Finally, many observables will be dominated by large logarithms, at
least in some range of the distributions. Combining fixed-order
calculations with a QED parton shower is a generic and powerful tool
to resum the leading logarithms. Thus, the mule might want to take a
shower after a hard day's work. The structure of FKS$^\ell$ is
particularly amicable to a YFS parton shower because it already
exploits the YFS structure. Initial (final) state collinear logarithms
can be resummed using the parton distribution (fragmentation function)
approach, which was recently extended to next-to-leading logarithmic
accuracy~\cite{Bertone:2019hks}. 

Apart from technical developments we have also made steps towards
being as open as possible with our results and facilitating their
cross checks. All data that has been used in the plots presented here
are available on a public git repository
\begin{lstlisting}[language=bash]
    (*@\url{https://gitlab.com/mule-tools/user-library}@*)
\end{lstlisting}
For each data set, we give the input data and a SHA1 identifier of the
code used to create it. Since the code is available as a Docker image,
anyone will be able to reproduce our results, regardless of operating
system and dependencies. We hope this will accelerate progress in the
theoretical description of low-energy particle physics experiments.

\subsection*{Acknowledgement} 

We are grateful to M.~Fael for help with implementing the vacuum
polarisation contributions as well as providing a MathLink interface
for {\tt alphaQED}. In addition we thank G.M.~Pruna for his
contributions to the implementation of the muon-decay matrix elements.
We also thank A.~Gurgone and N.~Schalch for their contributions to the
further development of \mcmule{}. Moreover, we acknowledge useful
discussions with A.~Antognini, V.~Ravindran, and M.~Spira.  It is a
pleasure to thank C.~Carloni Calame, M.~Chiesa, S.~Mehedi Hasan,
G.~Montagna, O.~Nicrosini, and F.~Piccinini for sharing their results
for validation of our muon-electron scattering results, as well as
R.~Bucoveanu and H.~Spiesberger for comparisons of electron-proton
scattering. We are further grateful to E.~Bagnaschi for his help
integrating {\sl udocker} into our workflow and to M.~Zoller for
validating some amplitudes with OpenLoops. TE and YU acknowledge
support by the Swiss National Science Foundation (SNF) under contract
200021\_178967.  YU further acknowledges partial support by a
Forschungskredit of the University of Zurich under contract number
FK-19-087. PB acknowledges support by the European Union's Horizon
2020 research and innovation programme under the Marie
Sk\l{}odowska-Curie grant agreement No 701647.

\clearpage
\begin{appendix}
  \section{Input parameters}
  \label{sec:appendix}

The computations in this paper are performed in the on-shell scheme
for the coupling and using pole masses. Accordingly, the input
parameters we use are~\cite{Tanabashi:2018oca}
\begin{align}
  \begin{split}
    \begin{aligned}
      \alpha&=1/137.035999084, &
      G_F&= 1.1663787\cdot 10^{-11}\, \MeV^{-2}, \\
      m_e&=0.510998950\, \MeV,  &
      m_\mu&=105.658375\, \MeV, \\
      m_\tau&=1776.86\, \MeV,  &
      m_p&= 938.272088\, \MeV\, .
\end{aligned}
\end{split}
\end{align}
To convert from MeV to $\rm \upmu b$ we use $(c\hbar)^2 =1=
3.89379372\cdot 10^{8}\, \MeV^{2}\rm \upmu b$. When presenting
branching ratios of the muon, we always divide by the full width,
determined from the lifetime $2.196981\cdot 10^{-6}\,\mbox{s}$ as 
$\Gamma_\mu = 2.995984\cdot 10^{-16}\, \MeV$.

The interaction of the photon of momentum $q = p' - p$ with the proton
is parametrised as
\begin{align}
 \bar u(m_p,p') \Big( 
        F_1(Q^2)\gamma^\mu+F_2(Q^2)\frac{{\rm i}\sigma^{\mu\nu}q_\nu}{2m_p}
      \Big)u(m_p,p) \,
\end{align}
where $Q^2=-q^2 \ge 0$. The form factors $F_1$ and $F_2$ are related
to the Sachs form factors as
\begin{align}
  G_E&= F_1 - \tau F_2,& G_M&= F_1+F_2,
\end{align}
where $\tau\equiv Q^2/(4m_p^2)$. Using the standard dipole
parametrisation with $\Lambda^2 = 0.71\,\GeV^2$ we set
\begin{align}
F_1(Q^2) = \frac{1+\kappa\tau}{1+\tau}\Big(1+\frac{Q^2}{\Lambda^2}\Big)^{-2}
\quad\text{and}\quad
F_2(Q^2) = \frac{-1+\kappa}{1+\tau}\Big(1+\frac{Q^2}{\Lambda^2}\Big)^{-2}
\,.
\end{align}
Here $\kappa=2.79284734$ is the proton's magnetic moment in units of the
nuclear magneton.

\end{appendix}

\bibliographystyle{JHEP}
\bibliography{mcmule.bib}

\end{document}